\documentclass[a4paper]{article}
\usepackage[totalwidth=480pt, totalheight=680pt]{geometry}

\usepackage{cite}
\usepackage{amssymb}
\usepackage{graphicx}
\usepackage{lipsum}
\usepackage{float}
\usepackage{booktabs}
\usepackage{multirow}
\usepackage{url}
\usepackage[table,xcdraw]{xcolor}
\usepackage{balance}
\usepackage{tikz}
\usetikzlibrary{shapes,arrows,positioning}
\usepackage[hidelinks]{hyperref}

\newcommand{\etal}{\emph{et al.}}

\title{A Roadmap Towards Resilient Internet of Things\\for Cyber-Physical Systems}
\author{%
  Denise~Ratasich~\textsuperscript{1},
  Faiq~Khalid~\textsuperscript{1},
  Florian~Geissler~\textsuperscript{2},
  Radu~Grosu~\textsuperscript{1},\\
  Muhammad~Shafique~\textsuperscript{1},
  Ezio~Bartocci~\textsuperscript{1}\\
\textsuperscript{1}~Institute of Computer Engineering, TU Wien, Austria\\(e-mail: firstname.lastname@tuwien.ac.at)
\\
\textsuperscript{2}~CPS Dependability Research Lab, Intel Corporation, Germany\\(e-mail: florian.geissler@intel.com)
}
\date{2018-10-29}

\begin{document}

\maketitle

\begin{abstract}
  The Internet of Things (IoT) is a ubiquitous system connecting many different
devices -- the \emph{things} -- which can be accessed from the distance.
The cyber-physical systems (CPS) monitor and control the things from the distance.
As a result, the concepts of dependability and security get deeply intertwined.
The increasing level of dynamicity, heterogeneity, and complexity
adds to the system's vulnerability, and challenges its ability to react to faults.
This paper summarizes state-of-the-art
of existing work on anomaly detection, fault-tolerance and self-healing,
and adds a number of other methods applicable to achieve resilience in an IoT.
We particularly focus on non-intrusive methods ensuring data integrity in the network.
Furthermore, this paper presents the main challenges in building a resilient IoT for CPS
which is crucial in the era of \emph{smart CPS} with enhanced connectivity
(an excellent example of such a system is connected autonomous vehicles).
It further summarizes our solutions, work-in-progress and future work to this topic
to enable ``Trustworthy IoT for CPS''.
Finally, this framework is illustrated on a selected use case:
A smart sensor infrastructure in the transport domain.

\end{abstract}

\section{Introduction}

\emph{Cyber-physical systems} (CPS)~\cite{Lee:2010,Ragunathan2010,Rajkumar2012,Ceccarelli:2016}
are the emerging smart information and communications technology (ICT)
that are deeply influencing our society in several application domains.
Examples include unmanned aerial vehicles (UAV), wireless sensor networks, (semi-) autonomous cars~\cite{Fagnant2015}, vehicular networks~\cite{Rajkumar2012} and a new generation of sophisticated 
life-critical and networked medical devices~\cite{Sokolsky2012}.
CPS consist of collaborative computational entities that are tightly 
interacting with physical components through sensors and actuators.  
They are usually federated as a system-of-systems communicating with 
each other and with the humans over the \emph{Internet of Things} (IoT), 
a network infrastructure enabling the interoperability of these devices.
\begin{figure}[h]
  \centering
  \includegraphics[width=\textwidth]{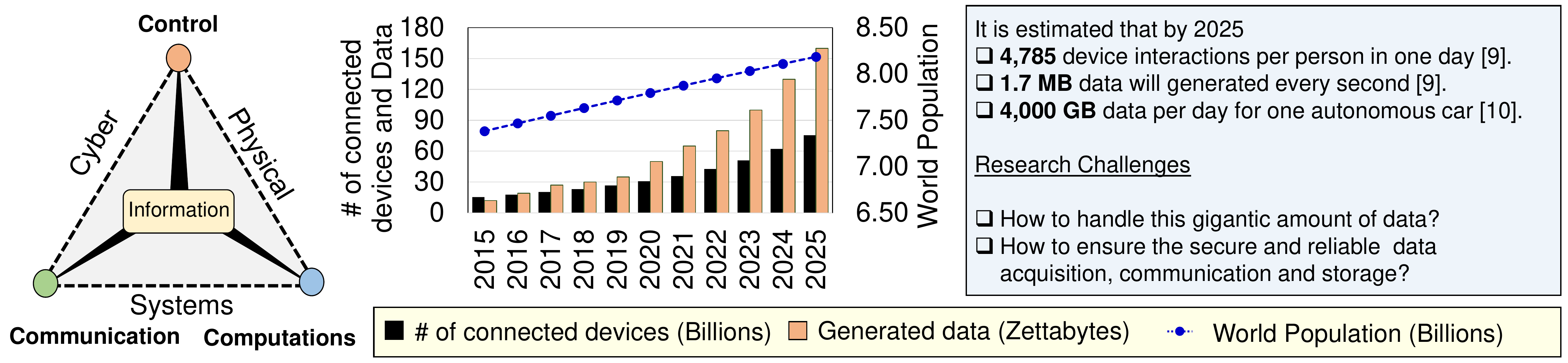}
  \caption{Introduction to CPS and Internet of Things (IoT) and growing trends of connected devices and generated data. Data Sources: \cite{statistica2017iot, statistica2018population,reinsel2017data, intel2018cars}}
  \label{fig:statistics}
\end{figure}

\subsection{Motivation for Resilient CPS}

The advent of the \emph{Internet} has revolutionized the communication between humans.
Similarly the CPS and IoT are reshaping the way in which we perceive and
interact with our physical world.  This comes at a price: these systems are 
becoming so pervasive in our daily life that failures and security vulnerabilities
can be the cause of fatal accidents, undermining their trustworthiness in the 
public eye.

Over the last years, popular mainstream newspapers have published several articles 
about CPS that are recalled from the market due to software and/or hardware bugs. 
For example in 2015, \emph{The New York Times} published the news~\cite{nytimes2015} about 
the finding of a software bug in Boeing 787 that could  cause  ``the plane power control 
units to shut down power generators if they were powered without interruption for 248 days". 
\emph{The Washington Post} has recently published an article~\cite{washingtonpost2018} 
about Fiat Chrysler Automobiles NV recalling over 4.8 million U.S. vehicles for a defect 
that prevents drivers from shutting off cruise control, placing them in a potential hazard.  
The recent accident of Uber's self-driving vehicle killing a pedestrian
shocked the world~\cite{guardian2018}, raising several concerns about 
the safety and trustworthiness of this technology.

With the connection of a CPS to the Internet, security becomes a crucial 
factor, too, that is intertwined with safety (``if it is not secure it is not
safe''~\cite{Bloomfield:2013}).  The tight interaction between the 
software and the physical components in CPS enables cyber-attacks 
to have catastrophic physical consequences.  \emph{The Guardian} reported last year~\cite{guardian2017} that over half a million pacemakers have been recalled by the 
American Food and Drug Administration due to fears that hackers could exploit cyber 
security flaws to deplete their batteries or to alter the patient's heartbeat.
In 2015 the BBC announced~\cite{bbc2016} that the black-out of the Ukraine power grid 
was the consequence of a malware installed on computer systems at 
power generation firms, enabling the hackers to get remote access to these computers.
In the same year two hackers have proved in front of the media~\cite{wired2015} that they  
could hijack a Jeep over the internet. 

The rise of the IoT, that is forecast to grow to 75 billions of devices in 2025~(Fig.~\ref{fig:statistics}),
is exacerbating the problem, by providing an incredibly powerful platform to 
amplify these cyber-attacks. 
An example is the MIRAI botnet that in 2016 have exploited more than 400000 devices 
connected through the IoT as a vehicle to launch some of the most potent 
distributed denial-of-service (DDoS) in history~\cite{Kolias2017}.

\begin{figure}[h]
  \centering
  \includegraphics[width=0.7\linewidth]{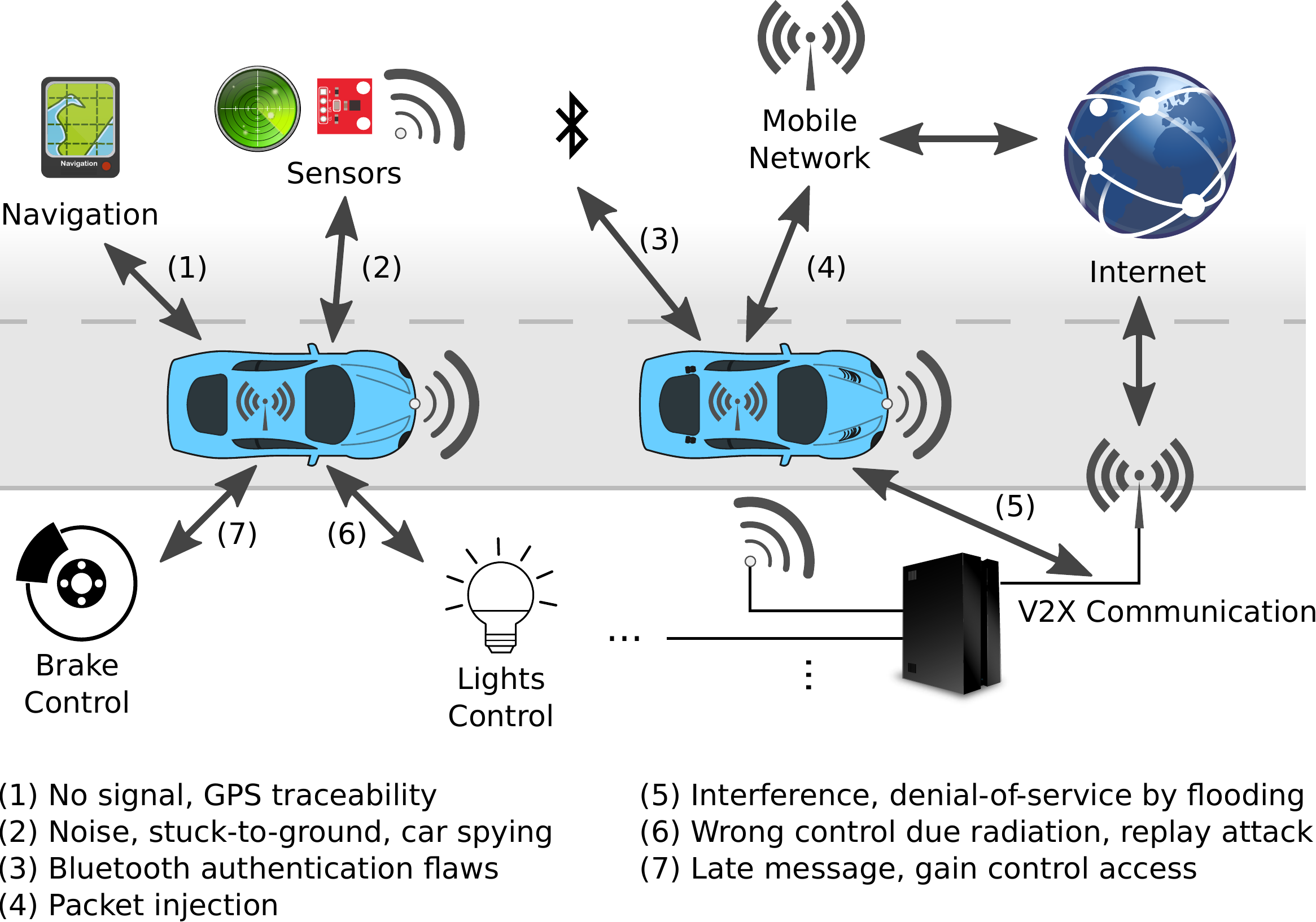}
  \caption{Examples of faults and threats in a connected vehicle.}
  \label{fig:fault-examples}
\end{figure}

Managing and monitoring such ultra large scale system is becoming extremely 
challenging. A desired property to achieve/enforce this
is to be \emph{resilient}, i.e., the service delivery (or functionality) that can 
justifiably be trusted persists, when facing
changes~\cite{Laprie:2008}. In other words, the system shall remain safe and
secure in the advent of faults and threats~(see Fig.~\ref{fig:fault-examples} for some examples in the automotive domain)
that could be even unpredictable at design time or could emerge during runtime~\cite{Laprie:2008,Bloomfield:2013}.

\subsection{State-of-the-Art}

Resilience has been identified and discussed as a challenge in
IoT~\cite{Atzori:2010,Vermesan:2011,Bloomfield:2013,wollschlaeger2017future}.
However, it has been mostly studied in other areas of computer science%
~(see Table~\ref{tab:sota_resilience}).
The majority of surveys focus on one building block of a resilient system, e.g., a CPS,
or one attribute of resilience.
For instance, some publications survey security by intrusion detection~\cite{butun2014survey,mitchell2014survey}
(e.g., based on machine learning / data mining~\cite{Buczak:2016}
or computational intelligence~\cite{Wu:2010}).
Recent surveys on the IoT~(Table~\ref{tab:sota_iot}) review definitions,
state IoT and research challenges or
discuss technologies to enable interoperability and management
of the IoT.
However, to the best of our knowledge,
resilience, adaptation and long-term dependability and security
have not yet been discussed in the context of IoT for CPS.

\newlength{\sotatabwidth}
\sotatabwidth0.18\linewidth
\begin{table*}[h]
  \centering
\resizebox{\textwidth}{!}{%
  \begin{tabular}{|p{0.17\textwidth}|p{\sotatabwidth}|p{\sotatabwidth}|p{\sotatabwidth}|p{\sotatabwidth}|p{\sotatabwidth}|p{\sotatabwidth}|p{\sotatabwidth}|}
    \hline
    \multirow{2}{0.17\textwidth}{\textsc{References}}
    & \multicolumn{3}{c|}{\textsc{Dependability Techniques}}
    & \multicolumn{2}{c|}{\textsc{Security Techniques}}
    & \multirow{2}{\sotatabwidth}{\textsc{Chal\-lenges}}
    & \multirow{2}{\sotatabwidth}{\textsc{Case Study}}
    \\ \cline{2-6}
    & \textsc{De\-tec\-tion} & \textsc{Di\-ag\-no\-sis} & \textsc{Re\-cov\-ery} & \textsc{De\-tec\-tion} & \textsc{Mit\-i\-ga\-tion} & & \\ \hline
    \hline
    \cite{Avizienis:2004,Laprie:2008,Kopetz:2011} & \checkmark (1) & & \checkmark (1) & & & \checkmark & \\ \hline
    \cite{lncs10457,Leucker:2009} & \checkmark & & & & & \checkmark & \\ \hline
    \cite{Chandola:2009} & \checkmark & & & \checkmark & & \checkmark & \\ \hline
    \cite{butun2014survey,mitchell2014survey,Buczak:2016} & & & & \checkmark & & \checkmark & \\ \hline
    \cite{Cheng:2009,Siva:2010,deLemos:2013,Weyns:2017,Kounev:2017} & & & \checkmark & & \checkmark & \checkmark & \\ \hline
    \cite{Papp:2016} & & & \checkmark & & \checkmark & \checkmark & \checkmark \\ \hline
    \cite{humayed2017cyber} & & & & \checkmark & \checkmark & \checkmark & \\ \hline
    \cite{Isermann:2006,Ghosh:2007,Psaier:2011,Elhady:2018} & \checkmark & \checkmark & \checkmark & & & \checkmark & \\ \hline
    This paper & \checkmark & \checkmark & \checkmark & \checkmark & \checkmark & \checkmark & \checkmark \\ \hline
  \end{tabular}%
}
  \caption{Comparison to resilience roadmaps and surveys (annotations: (1) terminology only).}
  \label{tab:sota_resilience}
\end{table*}
\newlength{\iottabwidth}
\iottabwidth0.22\textwidth
\begin{table*}[h]
  \centering
\resizebox{\textwidth}{!}{%
  \begin{tabular}{|p{0.17\textwidth}|p{\iottabwidth}|p{\iottabwidth}|p{\iottabwidth}|p{\iottabwidth}|p{\iottabwidth}|p{\iottabwidth}|}
    \hline
    \multirow{2}{0.17\textwidth}{\textsc{Ref\-er\-ences}}
    & \multirow{2}{\iottabwidth}{\textsc{En\-abling Technologies}}
    & \multicolumn{2}{c|}{\textsc{Resilience Techniques}}
    & \multirow{2}{\iottabwidth}{\textsc{Chal\-lenges}}
    & \multirow{2}{\iottabwidth}{\textsc{Ap\-pli\-ca\-tions}}
    & \multirow{2}{\iottabwidth}{\textsc{Case Study}}
    \\ \cline{3-4}
    & & \textsc{De\-pend\-abil\-i\-ty} & \textsc{Se\-cu\-ri\-ty} & & &
    \\ \hline
    \hline
    \cite{Atzori:2010,Vermesan:2011} & \checkmark & & & \checkmark & \checkmark & \\ \hline
    \cite{Zanella:2014} & \checkmark & & & \checkmark & & \checkmark \\ \hline
    \cite{wollschlaeger2017future} & \checkmark & \checkmark & & \checkmark & & \\ \hline
    \cite{Bloomfield:2013} & & & \checkmark & \checkmark & & \\ \hline
    \cite{Conoscenti:2016,ray2018survey,reyna2018blockchain} & \checkmark & & \checkmark & \checkmark & \checkmark & \\ \hline
    \cite{khan2018iot} & & & \checkmark & \checkmark & & \\ \hline
    \cite{sfar2018roadmap} & & & \checkmark & \checkmark & & \checkmark \\ \hline
    This paper & & \checkmark & \checkmark & \checkmark & & \checkmark \\ \hline
  \end{tabular}%
}
  \caption{Comparison to IoT roadmaps and surveys.}
  \label{tab:sota_iot}
\end{table*}

\subsection{Novel Contributions}

This paper provides an overview of the state-of-the-art to resilience
- that is dependability \emph{and} security - for the IoT.
We focus on resilience mechanisms that can be applied during runtime
and may be extended to adapt,
such that a system undergoing changes remains resilient.
We discuss a roadmap to achieve resilience,
and illustrate our recent work on this topic with a case study.
In particular:
\begin{itemize}
\item We summarize state-of-the-art methods and discuss recent work
  on detection, diagnosis, recovery and/or mitigation of faults.
  Due to the expected heterogeneous architecture, we specifically target non-intrusive methods
  which reason and act in the communication network or at the interfaces of the IoT devices.
\item We state the challenges of these techniques when applied in the IoT
  and depict a roadmap on how to achieve resilience in an IoT for CPSs.
\item Besides discussing several new perspectives,
  we further demonstrate some of our key methods/solutions and ongoing works
  on providing high resilience for the information collected and employed by the IoT
  in an automotive case study.
\end{itemize}

\subsection{Organization of the Paper}

\begin{figure}[t]
  \centering
  \includegraphics[width=\textwidth]{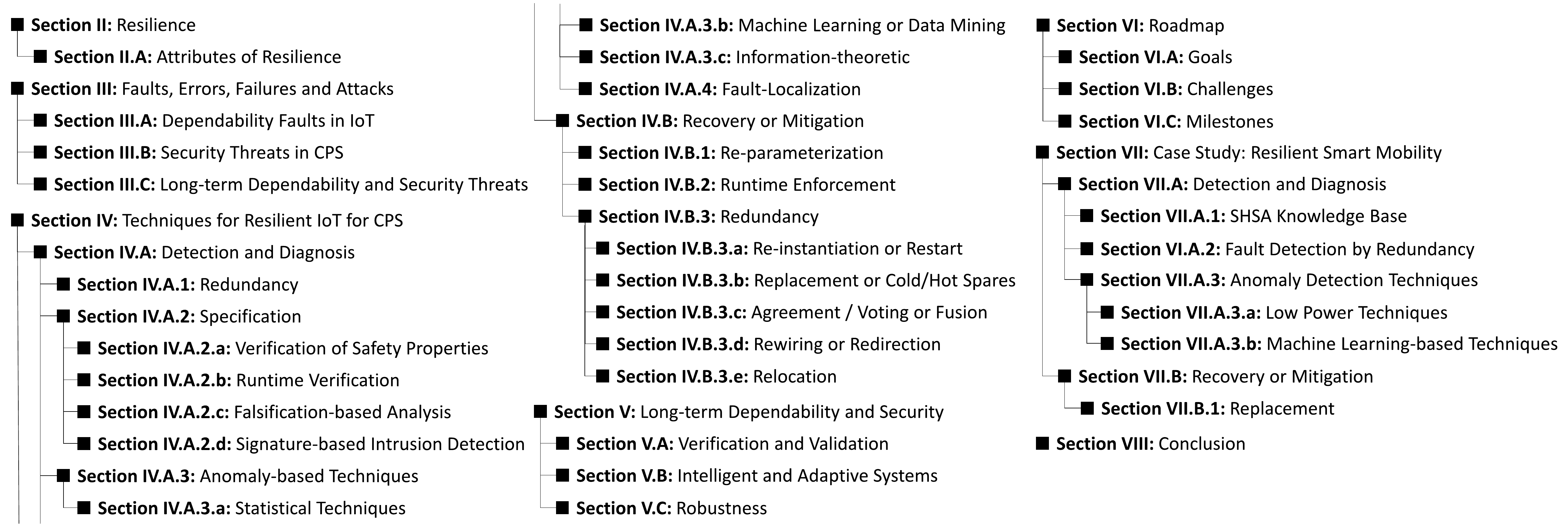}
  \caption{Organization of the paper.}
  \label{fig:organization}
\end{figure}

The rest of the paper is organized as follows (see also Fig.~\ref{fig:organization}).
The next two sections (Section~\ref{sec:resilience} and Section~\ref{sec:failures})
introduce the terminology around resilience,
fault types and examples,
building blocks of resilient systems and architectural layers
to the readers.
Section~\ref{sec:methods} collects state-of-the-art techniques
for fault detection and recovery.
Section~\ref{sec:long-term} states research challenges for resilience,
and particularly for the long-term dependability and security.
Section~\ref{sec:roadmap} discusses challenges and our roadmap to resilience in IoT
with several new perspectives.
Section~\ref{sec:casestudy} presents some of our key solutions to this topic
on the case study ``resilient smart mobility''.
Section~\ref{sec:conclusion} finally concludes the paper
with a discussion of the presented solutions and future work.

\section{Resilience}
\label{sec:resilience}

In order to provide a better understanding of resilient IoT,
we introduce resilience and its terminology in this section.

\subsection{Attributes of Resilience}

We desire the IoT for CPS to be dependable and secure throughout its entire life-cycle.
Avizienis~\cite{Avizienis:2004} defines the \emph{dependability} property of a system
to be the combination of following attributes: \emph{%
  availability (readiness for correct service),
  reliability (continuity of correct service),
  safety (absence of catastrophic consequences),
  integrity (absence of improper system alterations),
  maintainability (ability to undergo modifications and repairs)}.
Security includes availability, integrity and
\emph{confidentiality (the absence of unauthorized disclosure of information)}.

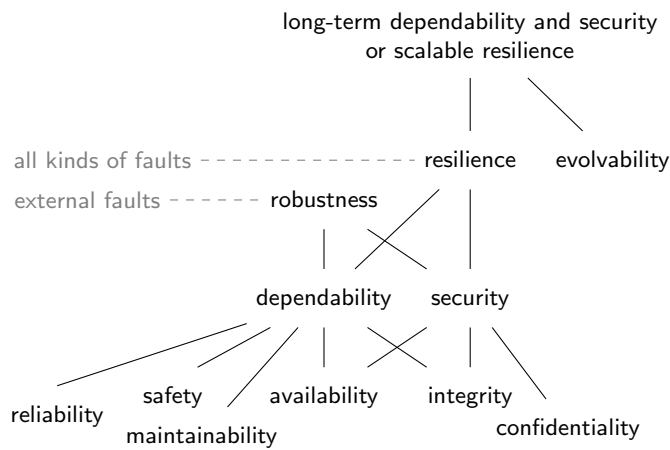
\begin{figure}[htb]
  \centering

\tikzstyle{block} = [rectangle, text height=1em, text depth=0.5em, text centered]
\tikzstyle{line2} = [text height=2em, text depth=1.5em]
\tikzstyle{maybe} = [draw=red]
\tikzstyle{note}  = [text=gray, text centered]


\begin{tikzpicture}[auto, step=1em, >=latex, font=\sffamily\small]
  \node [block] (r)  {resilience};
  \node [block, below       = 3em and 0em of r]   (sec)  {security};
  \node [block, below left  = 3em and 0.5em of r] (dep)  {dependability};
  \draw [-] (r) -- (dep);
  \draw [-] (r) -- (sec);

  \node [block, below       = 1.5em            of dep] (ds1)  {availability};
  \node [block, below       = 1.5em            of sec] (ds2)  {integrity};
  \node [block, below right = 2.7em and -1.2em of sec] (s3)   {confidentiality};
  \node [block, below left  =   3em and -1.5em of dep] (d1)   {maintainability};
  \node [block, left        = 1.8em            of ds1] (d2)   {safety};
  \node [block, below left  = 2.2em and 5em    of dep] (d3)   {reliability};
  \draw [-] (d1) -- (dep);
  \draw [-] (d2.50) -- (dep);
  \draw [-] (d3.north) -- (dep);
  \draw [-] (ds1) -- (dep);
  \draw [-] (ds2) -- (dep);
  \draw [-] (ds1) -- (sec);
  \draw [-] (ds2) -- (sec);
  \draw [-] (s3) -- (sec);

  \node [block, above = 1.5em of dep]   (rob)  {robustness};
  \draw [-] (dep) -- (rob);
  \draw [-] (sec) -- (rob);

  \node [note, left = 8.0em of r  ]   (af) {all kinds of faults};
  \node [note, left = 3.4em of rob]   (ef) {external faults};
  \draw [-, dashed, draw=gray] (af) -- (r);
  \draw [-, dashed, draw=gray] (ef) -- (rob);

  \node [block, line2, text width = 15em, above       = 2em of r  ]        (lt)  {long-term dependability and security\\ or scalable resilience};
  \node [block, below right = 2em and -5em of lt] (evo) {evolvability};
  \draw [-] (lt) -- (r);
  \draw [-] (lt) -- (evo);
\end{tikzpicture}
  \caption{Relation of system attributes in the context of resilience.}
  \label{fig:attributes}
\end{figure}

\emph{Robustness} can be considered as another attribute of dependability.
It has its roots in the control theory or CPS where
a system is called robust if it continues to function properly
under faults of stochastic nature (e.g., noise).
In recent work on the concepts of cyber-physical systems-of-systems~(CPSoS)~\cite{Ceccarelli:2016},
robustness is extended to consider also the security issues in CPS as well:
\emph{``Robustness is the dependability with respect to external faults
  (including malicious external actions)''}.
%
Figure~\ref{fig:attributes} summarizes the attributes of a resilient system.

A \emph{fault-tolerant} system recovers from faults to ensure the ongoing
service~\cite{Avizienis:2004}, i.e., achieving dependability and robustness of a system.

The term \emph{resilience} is often used by the security community
to describe the resistance to attacks (malicious faults).
Laprie~\cite{Laprie:2008} defines resilience
for a ubiquitous, large-scale, evolving system:
Resilience is \emph{``The  persistence  of  service  delivery  that  can justifiably be trusted, when facing changes.''}.
The author builds upon the definition of dependability
by giving the following short definition of resilience
\emph{``The  persistence  of  dependability  when  facing changes.''}.


A ubiquitous, heterogeneous, complex system-of-systems will typically change over time
raising the need for the dependability and security established during design time to scale up.
We therefore find the definition of resilience from Laprie~\cite{Laprie:2008}
a good fit to express the needs of an IoT for CPS.
A resilient IoT ensures the functionality when facing also unexpected failures.
Moreover, it should scale dependability and security when it comes to functional, environmental and technological changes~\cite{Laprie:2005} --
we refer this capability to as \emph{long-term dependability and security}.

However, to ensure resilience in a system, two important factors need to be analyzed:
\emph{i)} possible faults (the sources of dependability and security threats, see Sec.~\ref{sec:failures})
and
\emph{ii)} available detection and mitigation methodologies (techniques and actions to apply, see Sec.~\ref{sec:methods}).

\section{Faults, Errors, Failures and Attacks}
\label{sec:failures}

A \emph{failure} is an event that occurs when a system deviates from its
intended behavior. The failure manifests due to an unintended state - the error
- of one or more components of the system. The cause of an error is called the
fault~\cite{Avizienis:2004}.

\begin{table}[t]
  \centering
  \begin{tabular}{|l|p{0.7\linewidth}|}
    \hline
    \textsc{Fault Type} & \textsc{Examples} \\
    \hline \hline
    Physical
    & Broken connector (e.g., due to aging effects),
    radiation, noise, interference, power transients,
    power-down or short generated by an attacker, material theft (e.g.,
    copper), denial-of-service by jamming / signal interference \\
    \hline
    Development
    & Hardware production defect, hardware design error (``errata''), software
    bug in program or data (memory leaks, accumulation of round-off errors,
    wrong set of parameters), unforeseen circumstances (of the system and/or
    its environment), vulnerabilities,
    aging effects like memory bloating or leaking \\
    \hline
    Interaction
    & Input mistake, message collision, spoofing (obscure identity),
    modify information with a Trojan horse, no or late message
    delivery (e.g., by replay attack), denial-of-service by flooding (e.g.,
    bomb of connection requests), hacked sensor producing inaccurate or false
    data causing incorrect control decisions and actuator actions \\
    \hline \hline
    Permanent
    & Design faults, broken connector, noise, stuck-at ground voltage due to a
    short, logic bomb carried by a virus slowing down or crashing the system,
    aging effects (e.g., electromigration) \\
    \hline
    Transient
    & Radiation, power transients, input mistake, intrusion attempt (via
    vulnerabilities, e.g., heating the RAM to trigger memory errors) \\
    \hline
  \end{tabular}
  \caption{Main classifications of faults by~\cite{Avizienis:2004} with examples.}
  \label{tab:faults}
\end{table}

%
The source of a fault~(Table~\ref{tab:faults}) may be internal or external. Internal faults may be of
physical nature (e.g., broken component connector) or introduced by the design
(software/hardware bug). External faults originate from the environment (e.g.,
noise, radiation) or inputs (e.g., wrong or malicious usage of the system).
Faults can be mainly classified into transient and permanent faults. Although a
transient fault manifests only for a short period of time, it can cause an
error and might lead to a permanent failure. Physical faults
(internal/environmental) and inputs may be transient or permanent. Design
faults are always permanent.
Faults that cannot be systematically reproduced are often called
intermittent faults (e.g., effects of temperature on hardware,
a transient fault like a short in the circuit activated by a specific input).
Such faults lead to so-called soft errors.
A possible attack scenario (that is a malicious external fault) is often referred to as a \emph{(security) threat}.

\begin{figure*}[t]
  \centering
  \includegraphics[width=1\linewidth]{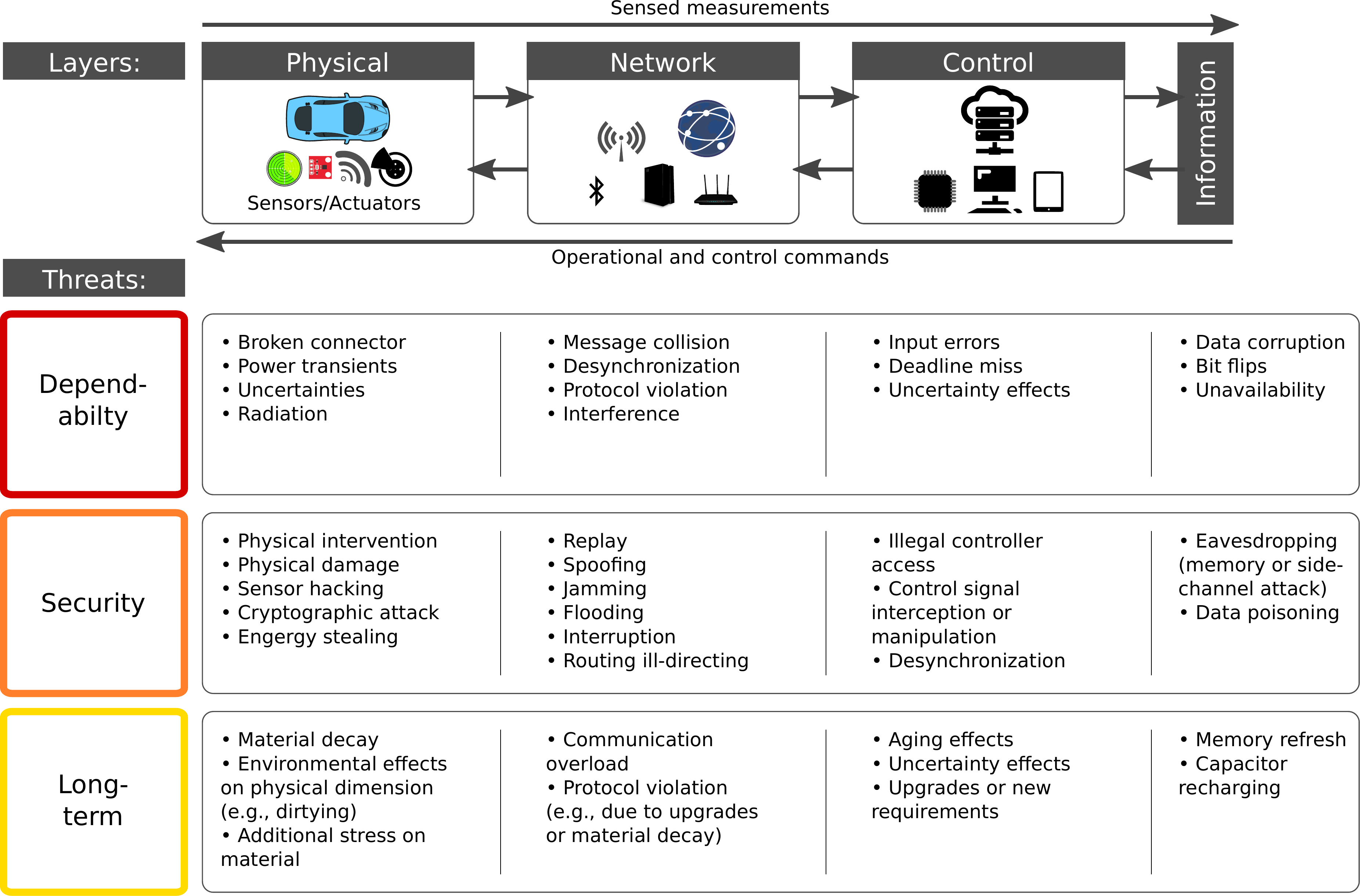}
  \caption{Dependability failures and security threats with respect to CPS layers.}
  \label{fig:CPS_arch}
\end{figure*}

Consider the CPS/IoT infrastructure shown in Figure~\ref{fig:CPS_arch}.
Faults
(e.g., radiation or a malicious signal for an actuator)
may occur at different layers of the architecture
(e.g., physical or control layer, respectively)%
~\cite{han2014intrusion}.
The physical layer is vulnerable to disruption, direct intervention or destruction of physical objects (e.g., sensors, actuators and mechanical components).
The network layer (here: the IoT) connects the devices.
The monitors and controllers in the control layer are vulnerable
to uncertainties of the environment and manipulation of measurements and control signals.
The information layer collects information and is particularly vulnerable to privacy and integrity issues.

The next two sections state examples of known
and emerging faults when the IoT meets CPS
(see also Fig.~\ref{fig:CPS_arch}).

\subsection{Dependability Faults in I\lowercase{o}T}
The IoT is susceptible to communication failures
particularly due to its size and heterogeneity.
Traditional CPS would avoid or mitigate such failures
by verification and sufficient testing of the design and final implementation
of the network component.
However, the IoT will evolve in technology and grow in size over time.
For instance, following faults may occur per CPS layer:
\begin{itemize}
\item \textbf{Physical Layer:}
  \begin{itemize}
  \item \textit{Interference:} Disruption of a signal.
    The number of connected devices and subsequently the radiation increases
    which may influence sensor measurements, transmitted messages or control signals~\cite{yaqoob2017internet}.
  \end{itemize}
\item \textbf{Network Layer:}
  \begin{itemize}
  \item \textit{Message Collision:}
    Similarly to interference, the number of communicating devices
    might trigger communication failures, e.g., collisions or an overload of the network.
  \item \textit{Protocol Violation:} Wrong message content
    due to different protocol version or protocol mismatch.
  \end{itemize}
\item \textbf{Control Layer:}
  \begin{itemize}
  \item \textit{Deadline Miss:} Late control signal reception.
    Control loops still have to follow the timing constraints of a CPS application.
  \item \textit{Misusage:} Send/set wrong inputs to a component, e.g.,
    due to wrong or incomplete syntactical and/or semantic information about the device.
  \end{itemize}
\item \textbf{Information Layer:}
  \begin{itemize}
  \item \textit{Unavailability:} Missing information caused by a technology update.
    Things might be connected, disconnected or updated in the IoT.
  \end{itemize}
\end{itemize}

\subsection{Security Threats in CPS}

Security has been a topic since the beginning of computer networks identifying vulnerabilities
(that is an internal fault or a weak point in a system enabling an attacker to alter the system~\cite{Avizienis:2004})
and avoiding or mitigating malicious attacks in devices.
However, in CPS additional vulnerabilities arise
given the connection to the physical domain and
the uncertain behavior of the physical environment~\cite{humayed2017cyber,shafique2018intelligent}.
For instance, following attacks may be applied per CPS layer:
\begin{itemize}
\item \textbf{Physical Layer:}~\cite{wurm2017introduction,myers2017automated,makedon2009event}
    \begin{itemize}
        \item \textit{Information Leakage}: Steal critical information from devices, e.g., secret keys or side channel parameters~\cite{zdancewic2001secure,xu2017security,chhetri2017fix,conti2018leaky,chhetri2018information}.
        \item \textit{Denial of Service (DoS):} Manipulate several parameters to perform a denial of service attack, e.g., hack the power distribution network to drain the energy \cite{kriebel2018robustness,balda2017cybersecurity}, destroy the sensors or actuators (in case of physical access), add extra power/communication load.
    \end{itemize}

\item \textbf{Network Layer:}
  W.r.t. security, this is the most vulnerable layer in a CPS
  because of the vast possibilities of attacks on communication networks which emerged over the years~\cite{mo2009secure,ali2018wsn,shoukry2013non,radcliffe2011hacking}.
    \begin{itemize}
        \item \textit{Jamming:} Overload the communication network by introducing fake traffic~\cite{adepu2017waterjam,li2015jamming,peng2018energy,peng2018optimal}.
        \item \textit{Collision:} Manipulate the timing, power and/or frequency of a network to trigger metastable states which eventually lead to data collision or violation of communication protocols ~\cite{zheng2017towards,cao2017probabilistic,wang2018data,dadras2018insider}.
        \item \textit{Routing ill-direct:} Manipulate the routing mechanism leading to data collision, data flooding and selective forwarding of data~\cite{hatzivasilis2017scotres,ali2018distributed}.
    \end{itemize}

\item \textbf{Control Layer:}
  \begin{itemize}
  \item \textit{Desynchronization:} Violate the timing or manipulate clocks~\cite{amrouch2017emerging,poudel2017design,lanotte2017formal,yampolskiy2018security}.
  This can also lead to a DoS~\cite{dong2017jamming} and/or information leakage~\cite{nichols2018hybrid,moore2017power,rekhis2017securing}.
  \end{itemize}

\item \textbf{Information Layer:}
  \begin{itemize}
  \item \textit{Eavesdropping:}
    Steal or sniff information. This is one of the major threats related to privacy.
  \end{itemize}
Moreover, information can also be manipulated to perform several attacks, i.e., jamming, collision or DoS.
\end{itemize}

The potential threats and consequences can be expressed in \emph{security threat models} for CPS~\cite{humayed2017cyber}.
To define a certain threat model, the following factors have to be identified:
\begin{enumerate}
    \item \textbf{Source/Attacker:} All the possible factors/actors which intentionally disturbs or interrupts the behavior or functionality of the CPS ~\cite{humayed2017cyber}. 
    \item \textbf{Attack Methodology}: The methodology or framework used to perform the attacks. However, it depends upon  the attacker's capabilities (available computational power, access to CPS resources and layers, etc.) motive (reason for the attack)~\cite{humayed2017cyber} and the type of attack vectors.  
    \item \textbf{Consequences/Payload}: The consequences of the actions that a successful attack performs to achieve its motive,
      e.g., compromising the confidentiality~\cite{aguma2018introduction},
      integrity~\cite{ahmed2017noisense},
      availability~\cite{gunduz2018reliability},
      privacy~\cite{fink2017security} and safety~\cite{marquis2018toward} of the CPS
      or information stealing~\cite{humayed2017cyber}.
\end{enumerate}
Table \ref{tab:sec_CPS} provides a summary of the possible threat models for each layer of CPS.
\begin{table*}[h]
	\centering
    \resizebox{1\linewidth}{!}{
\begin{tabular}{|l|l|l|l|l|l|l|}
\hline
\textsc{Layers} & \textsc{Physical} & \textsc{Sensor/Actuator} & \textsc{Communication} & \textsc{Control} & \textsc{Information} & \textsc{Integration Level} \\ \hline \hline
\textsc{Attackers} & M, D, E & M, D, E & M, D, E & M, D, E & M, D, E & M, D, E \\ \hline
\textsc{\begin{tabular}[c]{@{}l@{}}Methodology\end{tabular}} & \begin{tabular}[c]{@{}l@{}}Physical \\ Intervention\end{tabular} & \begin{tabular}[c]{@{}l@{}}Hacking, Control Access, \\ Data Manipulations\end{tabular} & \begin{tabular}[c]{@{}l@{}}Replay, Sybil, Jamming, \\ Flooding, Spoofing\end{tabular} & \begin{tabular}[c]{@{}l@{}}Control Access, \\ Eavesdropping\end{tabular} & Eavesdropping & \begin{tabular}[c]{@{}l@{}}All possible control \& \\ communication attacks\end{tabular} \\ \hline
\textsc{Payloads} & \begin{tabular}[c]{@{}l@{}}DoS, \\ Aging Reliability\end{tabular} & \begin{tabular}[c]{@{}l@{}}Energy Stealing, DoS, \\ Information Leakage, \\ Desynchronization\end{tabular} & \begin{tabular}[c]{@{}l@{}}Energy Stealing, DoS, \\ Information Leakage, \\ Desynchronization\end{tabular} & \begin{tabular}[c]{@{}l@{}}Information Leakage, \\ DoS, Desynchronization\end{tabular} & \begin{tabular}[c]{@{}l@{}}Information \\ Leakage\end{tabular} & \begin{tabular}[c]{@{}l@{}}Energy Stealing, DoS, \\ Information Leakage, \\ Desynchronization\end{tabular} \\ \hline
    \end{tabular}}
    \caption{Threat models for different CPS layers (M: Manufacturer, D: Designer, E: External Attacker).}
    \label{tab:sec_CPS}
\end{table*}

\subsection{Long-term Dependability and Security Threats}
\label{sec:long-term-threats}
The IoT and CPS will undergo changes over time,
especially when subjected to long operational duration
(over decades like in autonomous vehicles).
Following aspects of the change~\cite{Laprie:2005} might trigger faults~(see examples per CPS layer in Fig.~\ref{fig:CPS_arch}).
\begin{enumerate}
\item \textit{Environmental:}
  Uncertainty of the physical world.
  Decay and aging of material and components.
\item \textit{Functional:}
  Different and/or new applications and requirements.
  Dynamic system, i.e., connecting/disconnecting devices.
\item \textit{Technological:}
  Different and/or new components (e.g., maintenance, upgrades, demands), devices, interfaces or protocols.
  Unknown attacks (zero-day malware).
\end{enumerate}

\subsection{Fault Behavior}

A failure manifests in a wrong content or timing (early, late or no message at
all) of the intended service.
Components may contain an error detection mechanism and additionally suppress
wrong outputs. Such components are called \emph{fail-silent}. Some components
may automatically stop their execution on failures or halt crash, so-called
\emph{fail-stop} components. However, an erroneous component may provide wrong
outputs, i.e., the service is erratic (e.g., babbling) which can cause other
services to fail. In the worst case the behavior/output of the failed component
is inconsistent to different observers (Byzantine
failure)~\cite{Avizienis:2004,Kopetz:2011}.

\section{Techniques for Resilient IoT for CPS}
\label{sec:methods}

There are various online and offline approaches to achieve
resilience in a system. Developers may try to prevent faults (e.g., by an
appropriate design, encryption or consensus), tolerate faults (e.g., by switching to a
redundant component or another pre-defined configuration), remove/mitigate
faults (e.g., isolate faulty components to avoid the propagation of faults) or
forecast faults (e.g., to estimate the severity or consequences of a
fault)~\cite{Avizienis:2004}.
We want to focus on the possibilities to fulfill the following
requirements regarding resilience:
\begin{itemize}
\item \textbf{R1: Detection and identification of faulty, attacked or failed components during runtime in the IoT.}
  Faulty or already failed components shall be detected
  to be able to maintain or recover to a healthy system state
  providing correct system services.
\item \textbf{R2: Autonomously maintain resilience in the IoT.}
  Ensure the functionality of a dynamic and heterogeneous system in the presence of faults,
  i.e., recover from failures in an automatic fashion.
\end{itemize}

The following two sections give an overview about methods split
into \emph{detection and diagnosis}, and \emph{recovery or mitigation} of failures.
They summarize
background and terminology,
highly-cited surveys ($\ge$100 citations according to Google Scholar),
recent surveys ($\ge$2015),
recent approaches not part of surveys / additional work,
and examples~(see distribution in Table~\ref{tab:references})
given the keywords in Table~\ref{tab:keywords}.
Note that we tried to cite original publications
and no derivations of basic fault-tolerant techniques.

\begin{table}[h]
  \centering
  \begin{tabular}{|p{0.45\linewidth}|p{0.45\linewidth}|}
    \hline
    \textsc{Detection and Diagnosis} & \textsc{Recovery or Mitigation} \\ \hline
    \hline
    anomaly detection, fault detection/diagnosis,
    security in CPS, intrusion detection,
    runtime monitoring, runtime verification,
    self-awareness
    &
    self-healing, self-adaptation, software adaptation,
    runtime reconfiguration, fault-tolerance, fault recovery, threat mitigation
    dependability, resilience
    \\ \hline
  \end{tabular}
  \caption{Keywords used to find relevant research.}
  \label{tab:keywords}
\end{table}

\begin{table}[h]
  \centering
  \begin{tabular}{|p{0.2\linewidth}|p{0.325\linewidth}|p{0.325\linewidth}|}
    \hline
    \textsc{Type} & \textsc{Detection and Diagnosis} & \textsc{Recovery or Mitigation} \\
    \hline \hline
    Background / Terminology &
    \cite{Avizienis:2004} \cite{Isermann:2006} \cite{Chandola:2009} \cite{Leucker:2009} \cite{lncs10457}
    &
    \cite{Avizienis:2004} \cite{Laprie:2008} \cite{Cheng:2009} \cite{Kopetz:2011} \cite{Ceccarelli:2016} \cite{Kounev:2017} \cite{Weyns:2017}
    \\ \hline
    Highly-cited surveys &
    \cite{Isermann:2006} \cite{Chandola:2009} \cite{mitchell2014survey} \cite{butun2014survey}
    &
    \cite{Kephart:2003} \cite{Ghosh:2007} \cite{Cheng:2009} \cite{Psaier:2011}
    \\ \hline
    Recent surveys &
    \cite{lncs10457} \cite{Buczak:2016} \cite{humayed2017cyber}
    &
    \cite{Papp:2016} \cite{Elhady:2018} \cite{reyna2018blockchain} \cite{khan2018iot}
    \\ \hline
    Additional &
    \cite{Henzinger96} \cite{Leucker:2009} \cite{Bissmeyer:2012} \cite{fauri2017system} \cite{lncs10457}
    &
    \cite{Castro:2002} \cite{Falcone10} \cite{Conoscenti:2016} \cite{Meng:2018} \cite{FalconeMRS18}
    \\ \hline
  \end{tabular}
  \caption{Collection and distribution of basic work
      (apart from derived and optimized techniques) per publication type
      (ordered by publication year, ascending).}
  \label{tab:references}
\end{table}

\subsection{Detection and Diagnosis}
\label{sec:reason}

Anomaly detection is the process to identify an abnormal behavior or
pattern. The abnormal behavior or service failure (e.g., wrong state, wrong
message content) is caused by a fault~\cite{Avizienis:2004}, e.g., a random
failure, a design error or an intruder.
Though this definition probably complies with all fault detection mechanisms
listed in this section, the various communities use different keywords
depending on the application or type of the mechanism.
The related term \emph{monitoring} is used in the field of runtime verification
to refer to the \emph{act of observing and evaluating temporal
  behaviors}~\cite{lncs10457}.
In the security domain the phrase \emph{intrusion detection} is used for reasoning about threats.

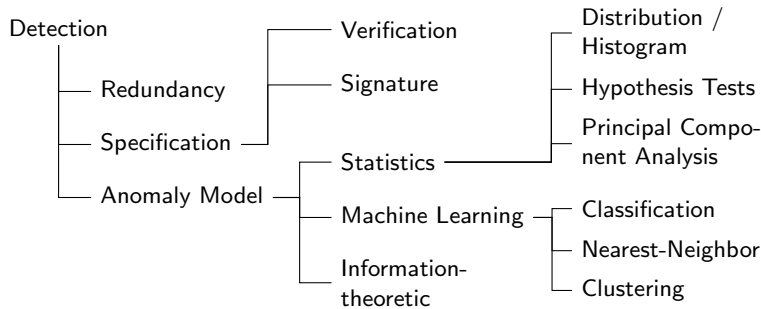
\begin{figure}[htb]
  \centering

\tikzstyle{block} = [rectangle, rounded corners, align=left]
\tikzstyle{maybe} = [draw=red]


\begin{tikzpicture}[auto, node distance=1em, >=latex, font=\sffamily\small]
  \node [] (reason) {Detection};
  \node [block, below right = 1em and -1em of reason] (t1)  {Redundancy};
  \node [block, below right = 3em and -1em of reason] (t2)  {Specification};
  \node [block, below right = 5em and -1em of reason] (t3)  {Anomaly Model};
  \draw [-] (reason) |- (t1);
  \draw [-] (reason) |- (t2);
  \draw [-] (reason) |- (t3);

  \node [block, below right = -5em and 9em of t2.west] (spec1) {Verification};
  \node [block, below right = -3em and 9em of t2.west] (spec2) {Signature};
  \draw [-] (t2.east) -| +(1em,0) |- (spec1.west);
  \draw [-] (t2.east) -| +(1em,0) |- (spec2.west);

  \node [block, below right = -2em and 9em of t3.west] (ad2) {Statistics};
  \node [block, below right =  0em and 9em of t3.west] (ad1) {Machine Learning};
  \node [block, below right =  2em and 9em of t3.west, text width=7em] (ad3) {Information-theoretic};
  \draw [-] (t3.east) -| +(1em,0) |- (ad1.west);
  \draw [-] (t3.east) -| +(1em,0) |- (ad2.west);
  \draw [-] (t3.east) -| +(1em,0) |- (ad3.west);

  \node [block, below right = -1.0em and 9em of ad1.west] (ml1) {Classification};
  \node [block, below right =  0.5em and 9em of ad1.west] (ml2) {Nearest-Neighbor};
  \node [block, below right =  2.0em and 9em of ad1.west] (ml3) {Clustering};
  \draw [-] (ad1.east) -| +(0.7em,0) |- (ml1.west);
  \draw [-] (ad1.east) -| +(0.7em,0) |- (ml2.west);
  \draw [-] (ad1.east) -| +(0.7em,0) |- (ml3.west);

  \node [block, below right = -6.2em and 9em of ad2.west, text width=7em] (st1) {Distribution / Histogram};
  \node [block, below right = -3.5em and 9em of ad2.west] (st2) {Hypothesis Tests};
  \node [block, below right = -2.0em and 9em of ad2.west, text width=7em] (st3) {Principal Component Analysis};
  \draw [-] (ad2.east) -| +(4em,0) |- (st1.west);
  \draw [-] (ad2.east) -| +(4em,0) |- (st2.west);
  \draw [-] (ad2.east) -| +(4em,0) |- (st3.west);
\end{tikzpicture}
  \caption{A taxonomy of methods for fault detection.}
  \label{fig:methods_reason}
\end{figure}

Halting failures (fail-stop or fail-silent behavior) can be detected by simple methods
like watchdogs or timeouts.
Faults that manifest in erratic or inconsistent values or timing need
a behavior specification, model or replica to compare against
(we therefore focus on these methods).
Such detection methods can be roughly separated w.r.t.
the knowledge used to compare to the actual behavior~(Fig.~\ref{fig:methods_reason}).

The expected or faulty behavior is represented either via
formal models or specifications (runtime verification)~\cite{Leucker:2009,lncs10457},
signatures describing attack behaviors~\cite{butun2014survey,mitchell2014survey},
learned models (classification, statistics)~\cite{Chandola:2009,Isermann:2006,butun2014survey,mitchell2014survey},
clusters or the data instances itself (nearest-neighbor)~\cite{Chandola:2009,mitchell2014survey}.

Another field of reasoning about failures is the root cause analysis or \emph{fault localization} which
identifies the reason why a fault occurs (e.g., a vulnerability of the system
or the first failed component which caused other components to fail due to
fault propagation).

\subsubsection{Redundancy}
Additional information sources can detect many types of
faults~\cite{Petit:2015}.
A simple method to verify a message's content or intermediate result is
plausibility checking or majority voting~\cite{Kopetz:2011}, e.g., by comparing a received
message's content against redundant information sources
(see also ``agreement'' in Sec.~\ref{sec:act}).
Nevertheless, redundancy is typically the last resort to increase the
resilience or to ensure a specific level of dependability because it is costly
when it is added explicitly (e.g., triple modular
redundancy often deployed in the avionics~\cite{Kopetz:2011}).

In hardware, fault detection by redundancy is also known as lockstep execution
where typically two computational units run the same operations in parallel to
detect faults~\cite{Poledna:1996,kriebel2016variability}. When three replicas are used, the fault can
be masked by majority voting (under the assumption that only one component can
fail at the same time), see also \emph{Triple~Modular~Redundancy~(TMR)} in Section~\ref{sec:act}.

However, some techniques exploit implicit or functional redundancy that is
already available in the system.
For instance, \cite{Bissmeyer:2012} combines anomaly detection with sensor
fusion. Their approach uses a particle filter fusing data of different sensors
and simultaneously calculating a value of trust of the information sources
derived from the normalization factor, i.e., the sum of weights of the
particles. When the weights of the particles are high, the information source
match the prediction and are rated trustworthy.
The authors in~\cite{Duerrwang:2017} propose to use hard-wired local data of an
automotive ECU to check the plausibility of a received control input.
Our method presented in Section~\ref{sec:casestudy} is based and relies upon
implicit (and explicit) redundancy too.

\subsubsection{Specification}

\begin{figure}[t]
  \centering
  \includegraphics[width=\textwidth]{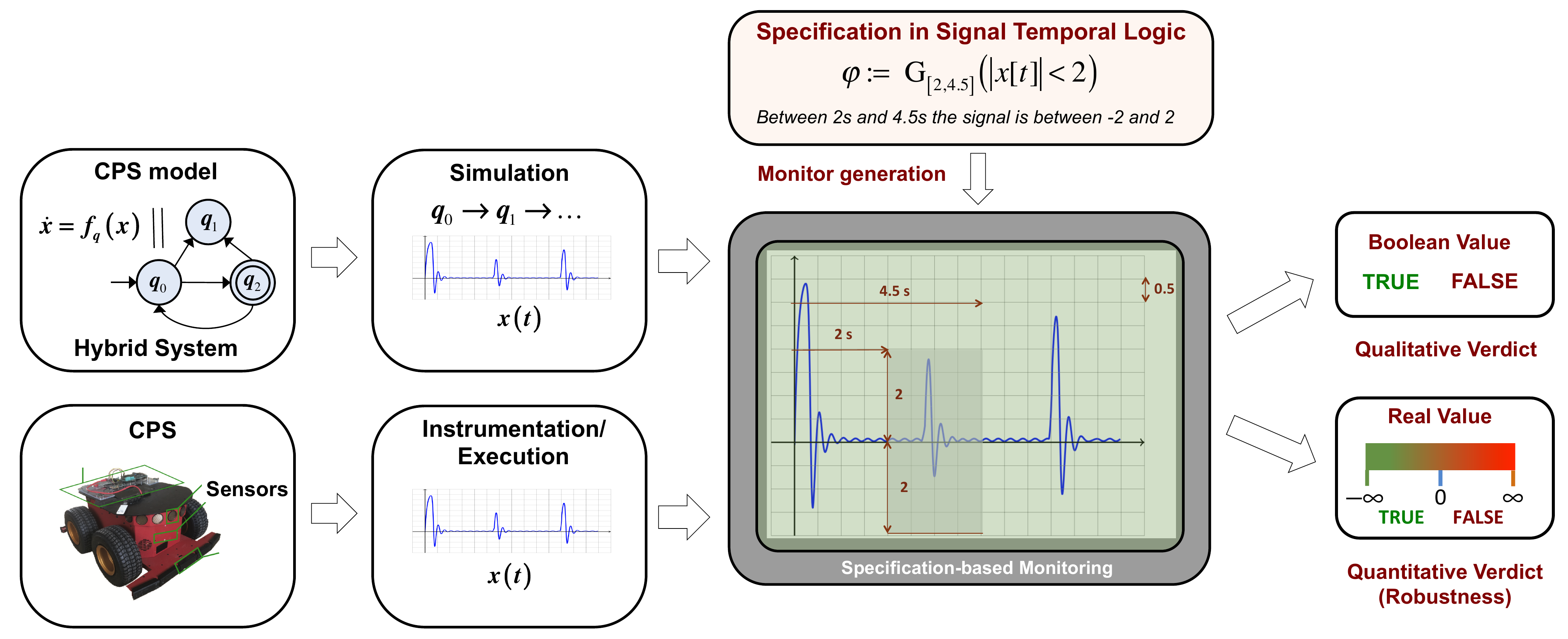}
  \caption{Specification-based monitoring can be employed either during the CPS execution or at design-time during the CPS model simulation.}
  \label{fig:monitoring}
\end{figure}

\paragraph{Verification of Safety Properties}
The IoT generally consists of spatially distributed and networked CPS. 
At design time, the CPS behavior can be modeled using hybrid systems, 
a mathematical framework that combines discrete transition systems 
capturing the computational behavior of the software component with 
continuous (often stochastic and nonlinear) ordinary differential 
equations (ODEs) describing the behavior of the physical substratum 
with which the software component is deeply intertwined.

Although there has been a great effort in literature to provide 
efficient computational techniques and tools~\cite{Althoff2013,Franzle2007,AsarinDM02,RayGDBBG15,chen2013flow,FrehseLGDCRLRGDM11,kong2015dreach,duggirala2015c2e2} to analyze safety properties in hybrid systems, 
the exhaustive verification (i.e., model checking) is in general undecidable~\cite{Henzinger96}.
The approaches currently available to check safety properties are based 
on generating conservative over-approximations of the state variables dynamics 
called \emph{flow pipes}~\cite{KongBH18} and on checking whether those intersect the unsafe regions of interest. However, these methods are generally limited to small scale CPS 
models. This limitation becomes more evident when we 
want to study more complex emergent behaviors, which result from 
the interactions among system components and that can be observed 
only by taking in consideration large scale CPS.

Hybrid systems are approximation models of the real CPS behavior 
and so their analysis may be not always faithful due to inevitable 
approximations errors (especially of the physical behavior) in 
the modeling phase.  Furthermore, CPS models are not always 
available for intellectual property issues and indeed CPS need 
to be studied as black box systems where we are not able 
to observe the internal behavior.

\paragraph{Runtime Verification}

A complementary approach to exhaustive verification is to equip 
CPS with monitors that verify the correctness of their execution.  
Monitoring consists of observing the evolution of the discrete and 
continuous variables characterizing the CPS behavior and deciding 
whether the observed trace of values is good or bad.  As 
Fig.~\ref{fig:monitoring} illustrates, these traces can be obtained 
by simulating the CPS design or can be observed during the CPS 
execution through the instrumentation of the system under test 
(SUT) (more details concerning instrumentation techniques can 
be found in~\cite{BartocciFFR18}).

Runtime verification (RV)~\cite{lncs10457} is a \emph{specification-based} 
monitoring technique that decides whether an observed trace of 
a SUT conforms to rigorous requirements written in a formal 
specification language.  The main idea of RV consists 
in providing efficient techniques and tools that enable the 
automatic generation of a software- or hardware-based monitor~\cite{fpga,SelyuninJNRHBNG17}
from a requirement. RV can provide useful information about 
the behavior of the monitored system, at the price of a 
limited execution coverage.

RV is nowadays a very well-established technique, widely employed 
in both academia and industry both before system deployment, for 
testing, verification, and post-deployment to ensure reliability, 
safety, robustness and security.

A typical example of formal specification language is the Linear 
Temporal Logic (LTL) introduced by Pnueli in~\cite{ltl}.  LTL 
provides a very concise and elegant logic-based language to specify 
sequences of Boolean propositions and their relations at different points 
in time.  LTL considers only the temporal order of the events and 
not the actual point in time at which they really occur.  For example, 
it is not possible to specify that a property should hold after 
one unit of time and before three and a half units of time.  

Real-time temporal logics~\cite{Alur1994}
overcome these limits by embedding a continuous time interval in 
the \emph{until} temporal operator.
Signal Temporal Logic~\cite{Maler2004,Donze2012} is a popular 
example of a real-time temporal logic suitable to reason about
the real-time requirements for CPS which has been proposed for detection 
of threats~\cite{jones2014anomaly}. 

Although reasoning about a single trace can provide an 
insight about safety properties, this is generally not sufficient 
to capture important information-flow security properties~\cite{Clarkson2010} 
such as noninterference, non-inference and information leakage. These
properties are called \emph{hyperproperties}, because in order 
to be verified, they need two or more execution traces of the system 
to be considered at the same time. 
In order to specify hyperproperties, both LTL and STL have 
been extended respectively in HyperLTL~\cite{ClarksonFKMRS14} 
and HyperSTL~\cite{Nguyen2017hyper} adding in the syntax both 
universal and existential quantifiers 
over a set of traces.  Runtime verification of such specification 
languages is still an open challenge (some preliminary 
results appeared in~\cite{BonakdarpourF16}), since the majority of the monitoring algorithms available are usually developed to handle 
only a single trace.

\paragraph{Falsification-based analysis and Parameter synthesis}

 As illustrated in Fig.\ref{fig:monitoring}, 
the Boolean semantics of STL decides whether a signal is correct or 
not with respect to a given specification.  However, this answer is 
not always informative enough to reason about the CPS behavior, 
since the continuous dynamics of these systems are expected to be 
tolerant with respect to the value of certain parameters, 
the initial conditions and the external inputs.

Several researchers have proposed to address this issue
by defining a {\em quantitative} semantics for STL~\cite{fainekos-robust,robust1}.
This semantics replaces the binary satisfaction relation with a 
quantitative {\em robustness degree} function that returns a real 
value (see Fig.\ref{fig:monitoring}) indicating how far is a signal from satisfying or violating 
a specification.  The positive and negative sign of the robustness 
value indicates whether the formula is satisfied or violated, respectively.

The notion of STL robustness was exploited in several tools~\cite{staliro,breach} for falsification analysis~\cite{DokhanchiEtal2015emsoft} and parameter synthesis~\cite{DonzeKR09hscc,BartocciBNS15} 
of CPS models.  On one hand, trying to minimize the robustness~\cite{staliro} 
is suitable to search counterexamples in the input space that 
violates (falsifies) the specification. 
On the other hand, maximizing the robustness~\cite{breach} can be used to tune 
the parameters of the system to improve its resilience.   To this end,
a global optimization engine is employed to systematically guide the search.

\paragraph{Signature-based Intrusion Detection}
Signature-based intrusion detection compares pre-defined behavior
(known as golden behavior or signature)
to identify the the abnormal event during runtime~\cite{butun2014survey}.
Though these techniques effectively identify the intrusion with a small number of false positives
they require a precisely calibrated signature~\cite{fauri2017system}.
Therefore, such techniques are not feasible if designers and IP providers are not trusted.
Such misuse-based intrusion detection typically cannot handle zero-day attacks that are new unknown attacks.
It is therefore often combined with anomaly detection (e.g., in~\cite{David:2015}).

\subsubsection{Anomaly-based Detection}

\paragraph{Statistical Techniques}
In statistical anomaly detection the data is fit into a statistical model. 
If a test instance occurs in the low probability region of the model, i.e., it is
unlikely to be generated by the model, then it is claimed to be an anomaly.
Statistical models can be specified with parameters when the underlying
distribution is known (e.g., is Gaussian). The parameters are trained
by machine learning~(ML) algorithms~\cite{Chandola:2009} or estimation~\cite{Isermann:2006}
describing the correct behavior of the system.
The inverse of the test instance's probability to be generated can directly be
used as anomaly score. Statistical tests can also be used to label or score a
test instance (e.g., box plot rule).

The model can be expressed by the data itself, e.g., in a histogram, by kernel
functions or particles, which is typically used when the distribution of the
data is unknown.
The test instances or samples may be evaluated by statistical hypothesis
tests. For instance, the Wilcoxon signed-rank test~\cite{Wilcoxon:1945}
compares two related samples to determine if they have the same underlying
distribution (which is unknown and does not have to be the normal distribution).

The principal component analysis (PCA) is used to project the data to lower
dimensions, i.e., it reduces the dimensionality of the data to a set of
uncorrelated variables. A test instance can be marked anomalous when the
projection on the components result in a high variance meaning that the test
instance does not fit the typical correlation of the data.

However, simple tests, Gaussian models and histograms are nowadays mostly
replaced by (deep) neural networks which stand out handling multivariate and
non-linear data.

\paragraph{Machine Learning or Data Mining}

Typical anomaly detection techniques based on machine learning
can be used with data where no domain knowledge is
available (e.g., black-box components like IP cores).
The models may be updated during operation. When the desired
behavior is known it can be expressed as formal model (specification-based
monitoring).

%
Classification-based anomaly detection learns a model (SVM, neural network,
Bayesian networks, rules or decision trees) given labeled training data (e.g.,
states and observations of the system) to cluster the test data into normal
classes and anomalies or outliers~\cite{Chandola:2009}.
Instead of labeling a test instance to a class, one may use scores representing
the likelihood of a test instance being an anomaly.
For instance, the authors in~\cite{Malhotra:2015} use recurrent neural networks
to detect anomalies in real-time data. The network models short and long term
patterns of time series and serves as a prediction model of the data. The error
between predicted and actual value serves as an anomaly score.

%
Nearest-neighbor-based detection techniques measure the distance from a data
instance under test to $k$ neighbors to identify anomalies. Different metrics
(e.g., euclidean distance) are applied to specify an anomaly score - that is
the likelihood of a data instance to be an anomaly. Another approach is to
measure the density that is the number of instances in the area specified by
the data instance under test given a radius.
The Nearest-Neighbor's complexity increases with the power of two of the number
of data instances. Unsupervised.

%
Data instances are first distributed into clusters (by clustering algorithms,
e.g., expectation maximization, k-means, self-organizing maps, many of which
use distance or density measures). An anomaly is a data instance that does not
fit into any cluster.

\paragraph{Information-theoretic}
By investigating the information content described by, e.g., the entropy of the
information, one may draw conclusion about anomalies in the data
(for information-theoretic measures characterizing regularity in data see~\cite{Lee:2001}).
When the entropy exceeds a threshold the test instance is marked as
anomaly. The threshold is defined by the set of anomalies. In highly irregular
data the gap between threshold and maximum entropy may be low (the set of true
anomalies is small).

\subsubsection{Fault-Localization}
When the fault detection only gives us the information about a 
failure happened in a subsystem, we need means to identify the 
exclusive part causing the failure.

This is often performed by root cause analysis~\cite{Wilson:1993}
or fault-localization~\cite{BartocciFMN18,LiuLNBB16,LiuLNBB16b,LiuLNB17,4344104,DeshmukhJMP18,WongGLAW16}.
In the software engineering community there is a considerable 
amount of literature about (semi-)automatic techniques assisting 
the developer to localize and to explain program bugs (for a 
comprehensive survey we refer the work in~\cite{WongGLAW16}).  
A well-established statistical approach, is the 
\emph{spectrum-based fault-localization} (SFL)~\cite{4344104},  
a technique that provides a ranking of the program components 
that are most likely responsible for the observed fault.

This approach has been employed recently also to localize faults 
in Simulink/Stateflow CPS models~\cite{LiuLNBB16,LiuLNBB16b,LiuLNB17,DeshmukhJMP18,BartocciFMN18}, 
displaying a similar accuracy with the same method applied 
to software systems~\cite{LiuLNBB16b}.
Although the classical SFL is agnostic to the nature of the oracle 
and only requires to know whether the system passes or not a 
specific test case, in~\cite{BartocciFMN18}, the authors have 
introduced a novel approach where the oracle is a specification-based
monitor. This enables to leverage the trace diagnostic method proposed in~\cite{FerrereMalerNickovic15} and to obtain more information 
(for example the segment of time where the  fault occurred) about 
the failed tests improving the fault-localization.

Often this approach is only applied offline for debugging processes, 
however, it can be used to isolate a failed HW/SW component from 
the system to avoid fault propagation or trigger its recovery.

\subsection{Recovery or Mitigation}
\label{sec:act}

\begin{figure}[htb]
  \centering

\tikzstyle{block} = [rectangle, rounded corners, align=left]
\tikzstyle{maybe} = [draw=red]


\begin{tikzpicture}[auto, node distance=1em, >=latex, font=\sffamily\small]
  \node [] (act) {Recovery or Mitigation};
  \node [block, below right = 1em and 0em of act] (act1) {Re-Parameterization};
  \node [block, below right = 3em and 0em of act] (act3) {Runtime Enforcement};
  \node [block, below right = 5em and 0em of act] (act2) {Redundancy};
  \draw [-] (act) |- (act1);
  \draw [-] (act) |- (act2);
  \draw [-] (act) |- (act3);

  \node [block, above right = 3em and -3em of act1] (par1) {Optimization};
  \node [block, above right = 1em and -3em of act1] (par2) {Rule-based};
  \draw [-] (act1.north) |- (par1.west);
  \draw [-] (act1.north) |- (par2.west);

  \node [block, below right = -9em and 8em of act2] (red1) {Re-Instantiation};
  \node [block, below right = -7em and 8em of act2] (red2) {Replacement};
  \node [block, below right = -5em and 8em of act2] (red3) {Agreement};
  \node [block, below right = -3em and 8em of act2] (red4) {Rewiring};
  \node [block, below right = -1em and 8em of act2] (red5) {Relocation};
  \draw [-] (act2.east) -| +(7em,0) |- (red1.west);
  \draw [-] (act2.east) -| +(7em,0) |- (red2.west);
  \draw [-] (act2.east) -| +(7em,0) |- (red3.west);
  \draw [-] (act2.east) -| +(7em,0) |- (red4.west);
  \draw [-] (act2.east) -| +(7em,0) |- (red5.west);
\end{tikzpicture}
  \caption{A taxonomy of methods for recovery or mitigation.}
  \label{fig:methods_act}
\end{figure}
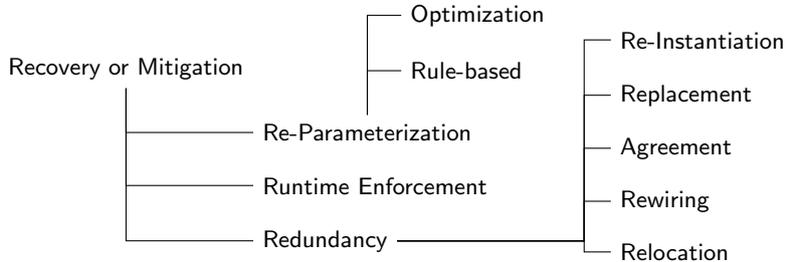

Broadly speaking, a system can be adapted by changing its parameters or its
structure (architecture)~\cite{Cheng:2009,Kounev:2017}.
Following four action types of possible re-configurations are defined by
\cite{Papp:2016} (splitting structural adaptation into further classes):
\emph{re-parameterization} to change the parameters of a component,
\emph{re-instantiation} to create and remove components, \emph{rewiring} to
redirect connections between components or \emph{relocation} to migrate
functionality to another platform.
The latter three action types require redundancy to some extent.
We extend and refine these types in the following (Fig.~\ref{fig:methods_act}).

Unless otherwise stated,
the adaptation can be applied on different architectural levels of the system.
For instance, the change of the clock speed or other hardware
parameters is the re-parameterization on the physical level of a
device. Changing the receiver of a software component's output is rewiring on the process/task level.

\subsubsection{Re-Parameterization}
In general, a re-parameterization (or reconfiguration) switches to another
configuration of one or more components that is typically no longer the optimal
setting, i.e., the quality of service is decreased (graceful degradation).
Adaptation of parameters requires knowledge about the underlying algorithm of the
erroneous component and is therefore typically performed by the component
itself or within a subsystem.
The configuration can be selected by optimization~\cite{Marler:2004},
or a reasoner
based on a set of rules, an ontology or a logic program~\cite{Papp:2016}.
Approaches from the control theory use state observers or estimators
to derive parameters to mitigate stochastic faults~\cite{Isermann:2006}.
For instance, an adaptive Kalman filter~(AKF)~\cite{Mehra:1970}
changes its filter parameters during runtime based on the inputs.
For instance, the measurement covariance can be increased
when an input signal gets worse or even permanently fails (cf.: a traditional KF or state
estimator mitigates noise and transient failures only).

\subsubsection{Runtime Enforcement}
Runtime enforcement~\cite{Falcone10,FalconeMRS18}  merges runtime verification with adaptation.  This powerful technique ensures that a program conforms to its specification.
A so-called enforcer acts on the interface of a component
changing inputs or outputs to comply with a set of formal properties.
The enforcer uses an automaton and/or rules to correct the IO in case of faults.
This approach has been pioneered by the work of Schneider~\cite{Schneider00} on security automata which halt the program whenever it deviates from a safety requirement.
Since then, there has been a great effort in the RV community 
to define new enforcement mechanisms with  primitives~\cite{LigattiBW05,FalconeMFR11,BielovaM12,DolzhenkoLR15,Ylies2016} or 
that support more expressive specifications~\cite{PinisettyFJMRN14,FalconeM15,RenardRF17}.

\subsubsection{Redundancy}
%
Redundant components ensure availability (passive) and increase reliability (active).
Failed components can be re-instantiated, replaced by spares, mitigated by voting or fusion, rewired or relocated~\cite{Psaier:2011,Papp:2016}.

\paragraph{Re-Instantiation or Restart} 
A straightforward fault-tolerance method is to restart a failed software
component. The tasks or the system typically saves checkpoints or output
messages of components on a periodic basis to roll back to a healthy
state~\cite{Meneses:2015}. The restart might be combined with a
re-parameterization.
Checkpointing/restart techniques are well studied for operating
systems~\cite{Sancho:2005} and may be applied on fog nodes or cloud servers.
The primary/backup approach activates a typically aperiodic backup task if the
primary task fails~\cite{Ghosh:1997}.
Adaptations in hardware and software also mitigate reliability threats
while considering the optimization cost constraints~\cite{li2013raster,li2013dhaser,li2013cser}.
Similarly, runtime reconfiguration polices have also been proposed to mitigate the reliability threats in microprocessors~\cite{li2017fine}.

\paragraph{Replacement or Cold/Hot Spares}
The simplex architecture~\cite{Seto98} considers two redundant
subsystems. A highly dependable subsystem jumps in when the high-performance
subsystem fails.
Triple modular redundancy~(TMR) replicates HW and/or SW components to mask
failures (through a voter, i.e., includes detection). The replicates are in the
best (but most costly) case diverse w.r.t. their design such that also design
and input errors can be masked~\cite{Kopetz:2011}.
Such hardware redundancy is typically added during design time
and used in closed, non-elastic systems.
To exploit these techniques, several reliability resilient microprocessor designs \cite{shafique2014dark,henkel2014multi,henkel2013reliable} and corresponding software layer controls \cite{rehman2014dtune,rehman2011reliable,rehman2014reliability} have been proposed to ensure the resilience towards reliability threats, i.e., soft errors.
Typically, TMR-based solutions possess a large area and power overhead.
However, adaptive-TMR solutions~\cite{kriebel2014aser} can trade-off between power budget and reliability threats.
Similarly, software and hardware error masking techniques \cite{shafique2013exploiting,henkel2014multi} exploit the dark silicon (under-utilized areas) in multi-core systems \cite{gnad2015hayat} to mitigate faults.
However, an IoT orchestrator can maintain a directory of available services
and redirect resource requests if necessary.

Implicit redundancy like related observations in a system (in contrast to
traditional redundancy that is the explicit replication of components) can be
exploited by structural adaptation. A substitute component is instantiated to
replace the failed component which includes also rewiring and possibly also a
relocation~\cite{Hoeftberger:2015,Ratasich:2018} (see Sec.~\ref{sec:casestudy}
for an example).

\paragraph{Agreement / Voting or Fusion}
Byzantine failures (inconsistent failures to different observers)
typically caused by malicious attacks can be detected and tolerated using replicas
(here: redundant services on different nodes of a distributed system)
by agreement or consensus on the outputs~\cite{Castro:2002}.
The output of redundant components can be combined or fused,
e.g., via filters or fuzzy logic~\cite{Elhady:2018}.
However, through recent implementations and usage in cryptocurrencies~\cite{Nakamoto:2008,Miller:2014}
the attention is shifted towards smart contracts and blockchains
which ensure authentication and integrity of data~\cite{Conoscenti:2016,khan2018iot,reyna2018blockchain,Meng:2018}.
Basically, a blockchain is a series of data records
each attached by a cryptographically secure hash function
which makes it computationally infeasible to alter the blockchain.
However, blockchains suffer from complexity, energy consumption and latency and
therefore currently cannot be used for real-time anomaly detection or applied
by simple nodes with low computational power and restricted battery power budgets~\cite{Meng:2018}.
However, it is already examined to manage access to data (authorization),
purchase devices or computing power or manage public-key infrastructure in the IoT~\cite{Conoscenti:2016,Dorri:2017,alphand2018iotchain}.

\paragraph{Rewiring or Redirection} 
Broken links in mesh networks are typically reconfigured using graph theory
considering node properties and application requirements~\cite{Alcaraz:2016}.
A software component may route the task flow to a recovery routine~\cite{Psaier:2011}.
%

\paragraph{Relocation} 
Migration of software components or tasks are studied in the field of resource
optimization, utilization and dynamic scheduling on (virtual)
machines. Optimization algorithms~\cite{Marler:2004}, multi-agent
systems~\cite{Khamphanchai:2011} or reinforcement learning~\cite{Yan:2016} find
a new task configuration utilizing resources in case of a platform failure.
Tasks may also be migrated in advance when the health state
decreases~\cite{Meneses:2015}.
Cloud applications boost and emerge new technologies
like containerization, resource-centric architectures and microservices
which ease service orchestration in complex and elastic systems.
Dragoni \etal{}~\cite{Dragoni:2017microservices} prognoses increased dependability
using microservices which focus on small, independent and scalable function units
(cf. fault containment units in Kopetz~\cite{Kopetz:2011}), however, security
remains a concern.

\section{Long-Term Dependability and Security}
\label{sec:long-term}

During design time only a subset of failures and threats can be considered,
however, the changes of the system itself or the environment
can not be predicted which may lead to new possible fault scenarios. Moreover, over the period of time
(especially when considering systems deployed for several decades like autonomous vehicles),
new attacks can emerge
(adversarial machine learning, though, ML is decades old theory),
new vulnerabilities in the system can be unleashed
(some recent examples are Spectre and Meltdown in decades old technology of high-end processors),
and attackers may get more powerful and intelligent (e.g., learning based attacks).

We therefore believe that the IoT needs enhanced self-adaptation techniques (may be cognitive in nature) to achieve
long-term dependability and security.
For instance, apart from traditional fault-tolerance
like backup hardware/software components or checkpointing and restarting,
\emph{self-healing} is a promising approach
which is related to self-adaptation and self-awareness.
Self-aware systems learn the models of the system itself and its environment to
reason and act (e.g., self-healing) in accordance to higher-level goals (e.g.,
availability)~\cite{Kounev:2017}. The key feature of \mbox{self-*} or \mbox{self-X} techniques is
continuous learning and optimization
which is performed during runtime to evolve the models upon system changes.

To design and build a long-term dependable and secure IoT of smart CPS,
the following research questions need to be addressed first:
\begin{enumerate}
\item[1)] How to detect and separate subsystem failures and
  minimize the failure dependencies of the subsystems?
  How to guarantee the resilience of the system
  when applying machine learning and/or self-adaptation?
\item[2)]
  How to detect and recover compromised components
  with minimal performance and energy overhead?
  How to learn from unknown attacks on-the-fly and
  devise appropriate mitigation strategies online,
  e.g., online on-demand isolation, new fail-safe modes, etc. besides investigating fast learning of on-going attacks to minimize the attack surface?
\item[3)] How to ensure the robustness of the resilience mechanisms itself?
\end{enumerate}
To address these challenges, following techniques are envisioned to ensure long-term dependability and security.

\subsection{Verification and validation}
Ensuring the complex dependencies and integrity of several components
and subsystems within a system is a very challenging research question.
The state-of-the-art on dependability and security assurance is based on
model-driven design that consists of specifying rigorously the
structure and the behavior of the systems using formal models. 
These models are amenable to formal verification techniques~\cite{bohrer2018veriphy,mohsin2016iotsat,mohsin2017iotriskanalyzer,kang2018model} that can provide 
comprehensive guarantees about correctness of the system's 
properties. The accuracy of these models and
the test coverage limit the validity of the assurance.

The addition of data-driven learning-enabled subsystems 
introduces uncertainty in the overall design process and 
may result in an unpredictable emergent behavior.
This is because the operational behavior of these subsystems 
is a function of the data they train upon and it is very 
difficult to predict. 

This lack of predictability force to think novel 
approaches for ensuring long-term dependability and 
security. Here we could envision at least two possible 
interesting research directions to pursue.

One idea could be to take inspiration by the natural immune systems that protect
animals from dangerous foreign pathogens (i.e., bacteria, viruses, parasites).
In our case, we can think to have a specialized subsystem
that learns both how the surrounding environment evolves
and how to best react to attacks.  However, this approach
would leave the system vulnerable during the learning process. 

Another possible direction is to provide mechanisms to enforce
dynamic assurance of security and dependability at runtime. 
A similar approach in control theory can be found in the \emph{Simplex
Architecture}~\cite{Seto98} (SA). SA consists of a plant and two
version of the controllers: a pre-certified baseline
controller and a not certified high-performance controller.
A decision module decides whether to switch between the
two controllers depending on how much close is the 
high-performance controller to violate the safety region.
In our case, we could envision a dedicated decision module
(for example a runtime monitor) enabling a certain degree 
of autonomy and trust to its subsystems depending on how 
much the overall system is far to violate a certain safe 
and secure operating conditions.


\subsection{Intelligent and Adaptive Systems}

Long-term dependability and security can be achieved by intelligent and adaptive systems
(i.e., so-called smart or cognitive systems)
that consider uncertainties and changes throughout the lifecycle,
and increase their inherent robustness levels on-the-fly
autonomously through continuous self-optimization and self-healing.

Many of the techniques presented in the last section
do not handle dynamic systems or systems that evolve over time.
However, the approaches can be evaluated and extended by artificial intelligence,
or new techniques developed to cope with the elasticity of the IoT.

Machine learning-based fault detection and recovery are replacing traditional (pre-configured) techniques
because of their ability to extract new and hidden features from the complex and enormous amount of data \cite{munir2018design,park2018study}.
To design intelligent and adaptive ML-based secure sub/-systems,
first a trained model must be acquired with respect to safe, secure and dependable behavior
while considering uncertainties, unforeseen threats and failures, and design constraints.
Next, the trained model is integrated within these sub/-systems for online threat
and fault detection under the area and power constraints.
However, ML-based techniques typically do not consider
the limited computational resources, complexity, probably poor interoperability or real-time constraints of the IoT.
Therefore, one has to apply scalable and/or distributed techniques.

\subsection{Robustness}

The subsystem ensuring resilience shall be robust against the elasticity or dynamicity of the system.
The machine learning models need to be updated or reconstructed from the basic building blocks on system changes.
Models like deep neural networks (DNNs) and recurrent neural networks (RNNs)
are effective in classifying real-world inputs when trained over large data sets.
Unfortunately, the decision-making systems using NNs cannot be analyzed and rectified
due to the currently used black-box models of NNs.

However, AI systems used in industry (in particular safety-critical CPS) need to follow the strict regulations
and are expected to explain the reasoning behind their decision-making
which is not viable when using ML-based systems \cite{ching2018opportunities}.
Recent advances in AI, e.g., biologically inspired NNs,
may provide the necessary information to get certified.
For instance, Hasani~and~Lechner~\etal~\cite{hasani2018re}
are able to interpret the purpose of individual neurons
and can provide bounds on the dynamics of the NN.

Moreover, a NN has several security and reliability vulnerabilities w.r.t. data, e.g.,
data poisoning, model stealing or adversarial examples~\cite{hanif2018robust,kriebel2018robustness}.
To ensure the robustness in such NN-based decision system, several countermeasure have been proposed.
A common approach is to encrypt the data or underlying model~\cite{hanzlik2018mlcapsule,hynes2018efficient,Riazi2018deep}.
However, encryption works as long as the encryption techniques and confidence vectors remain hidden from the adversary. Moreover, it requires additional computational resources for encryption and decryption.
Other approaches are, e.g., watermarking \cite{chen2018deepmarks,rouhani2018deepsigns}, input transformation \cite{zantedeschi2017efficient} and adversarial learning \cite{rakin2018defend,papernot2015distillation}.
Note that these countermeasures protect the NNs against known attacks only.
Therefore, to ensure the robustness also under unknown attacks and unforeseen circumstances
formal verification-based approaches \cite{xiang2018verification,xiang2018reachability} are emerging as an alternate solution.

\section{Roadmap}
\label{sec:roadmap}

We are investigating techniques for anomaly detection
and self-healing to ensure resilience in IoT for CPS.

\subsection{Goals}

The overarching goal of our research is
to provide guidelines, methods and tools to enable a safe and secure IoT for CPS.

Our contributions are two-fold.
We increase the dependability of the IoT (and in further consequence, the CPSs using it) by self-healing
and the security by developing (semi-) automatic configuration, testing and threat detection.
%
We plan to address the following research questions:
\begin{itemize}
\item How to improve the resilience of the IoT by fail-operational mechanisms?
\item How to verify and monitor IoT components?
\item How to detect anomalies in the IoT with minimum performance and energy overhead?
\item How to ensure high resilience even under unpredictable attack and failure scenarios?
\item What architectural requirements are necessary to ensure resilience with these mechanisms?
\end{itemize}

In summary, the key research goals of our contribution are:
\begin{itemize}
\item Propose novel design methodologies and architectures for scalable resilience in IoT for CPS.
\item Propose an energy-efficient analysis (verification) and threat detection.
\item Propose a framework to design low power and ML-based run-time anomaly.
\item Propose a methodology to identify and assert the runtime safety and security properties.
\item Propose a self-healing mechanism for the IoT.
\end{itemize}

\subsection{Challenges}
\label{sec:challenges}

The resilience of systems using anomaly-based detection and self-healing
raise the following research questions and challenges.

\begin{itemize}
\item \textbf{C1: Resource Limitations.}
  The majority of IoT components are resource-constrained devices.
  The developer often has to trade off power, time and costs against resilience.
  Typically, small IoT devices
  like commercial off-the-shelf (COTS) microcontrollers may provide
  insufficient capabilities.
  Some technologies might therefore need hardware implementations (e.g., RV monitor)
  or should be designed as a lightweight and fully distributed,
  layered,
  or clustered service (e.g., a monitor per subsystem).

  For instance, one major challenge in anomaly detection
  is the data acquisition under the consideration of power and design constraints.
  This raises following research questions:
\begin{enumerate}
\item How to extract/acquire and analyze a particular characteristics during run-time while considering the design and power constraints?
\item How to reduce the area and energy overhead of the data acquisition, i.e., power-ports, for runtime measurement and modeling?
\end{enumerate}

\item \textbf{C2-1: Interoperability and Complexity.}
  The IoT is a large dynamic network of heterogeneous components.
  In particular, COTS or components protected by intellectual property (IP)
  may not provide a proper specification of its behavior
  for some of the detection and adaptation methods.
  Furthermore, new devices or subsystems may introduce unknown interfaces
  (here: unknown to the resilience-enabling technologies).
  In particular, this raises following research questions:
  \begin{enumerate}
  \item How to identify the reference communication behavior without any reference system?
  \item How to model the communication behavior which can be used to identify the anomalous behavior?
  \end{enumerate}
  In anomaly detection, for instance, one of the major challenge
  is to identify the appropriate golden/reference behavior
  which can be used to compare with online/offline behavior.
  This raises following particular research questions:
\begin{enumerate}
\item How to model/identify the reference/golden behavior
  that covers the key characteristics and can be scalable?
\item How to obtain the labeled data for supervised training to extract the reference model?
\item Which modeling techniques and corresponding characteristics are appropriate to identify the anomalous behavior with complete coverage?
\end{enumerate}

\item \textbf{C2-2: Interoperability and Sharing.}
  The devices of a CPS are specified during design time
  having a specific application in mind.
  The things of an IoT will most likely be shared between applications while
  different fog/cloud applications might request different QoS of the devices,
  e.g., regarding dependability.
  The methods therefore must also consider and combine
  the requirements of different applications and
  the value of trust of the information (e.g., used to derive actions).
  Due to the vast size of an IoT, a central mechanism most likely will not be able to
  cope with all the input data necessary to achieve resilience (considering memory and time constraints).

\item \textbf{C3: Real-Time and Scalability.}
  One major shift from sensor networks to the IoT
  is the control and manipulation of actuators from the distance,
  i.e., the IoT comprises a cyber-physical system.
  The CPS typically has to satisfy \emph{time constraints} (rates, deadlines)
  in order to function correctly.
  In such real-time applications the probing of information by a monitor
  or changes in the system (e.g., connection of new things, updates, recovery)
  shall not influence the timing behavior of the CPS.
  Furthermore, the timeliness to detect and react to critical failures
  has to be considered.
  %

\end{itemize}

However, the complexity and dynamicity of the network will leave the door ajar for some faults, e.g., physical faults, design errors or zero-day malware. Therefore a proper \emph{never-give-up} strategy~\cite{Kopetz:2011} to cope with unconsidered failures has to be developed.

\subsection{Milestones}

\begin{figure}[t]
  \centering
  \includegraphics[width=\textwidth]{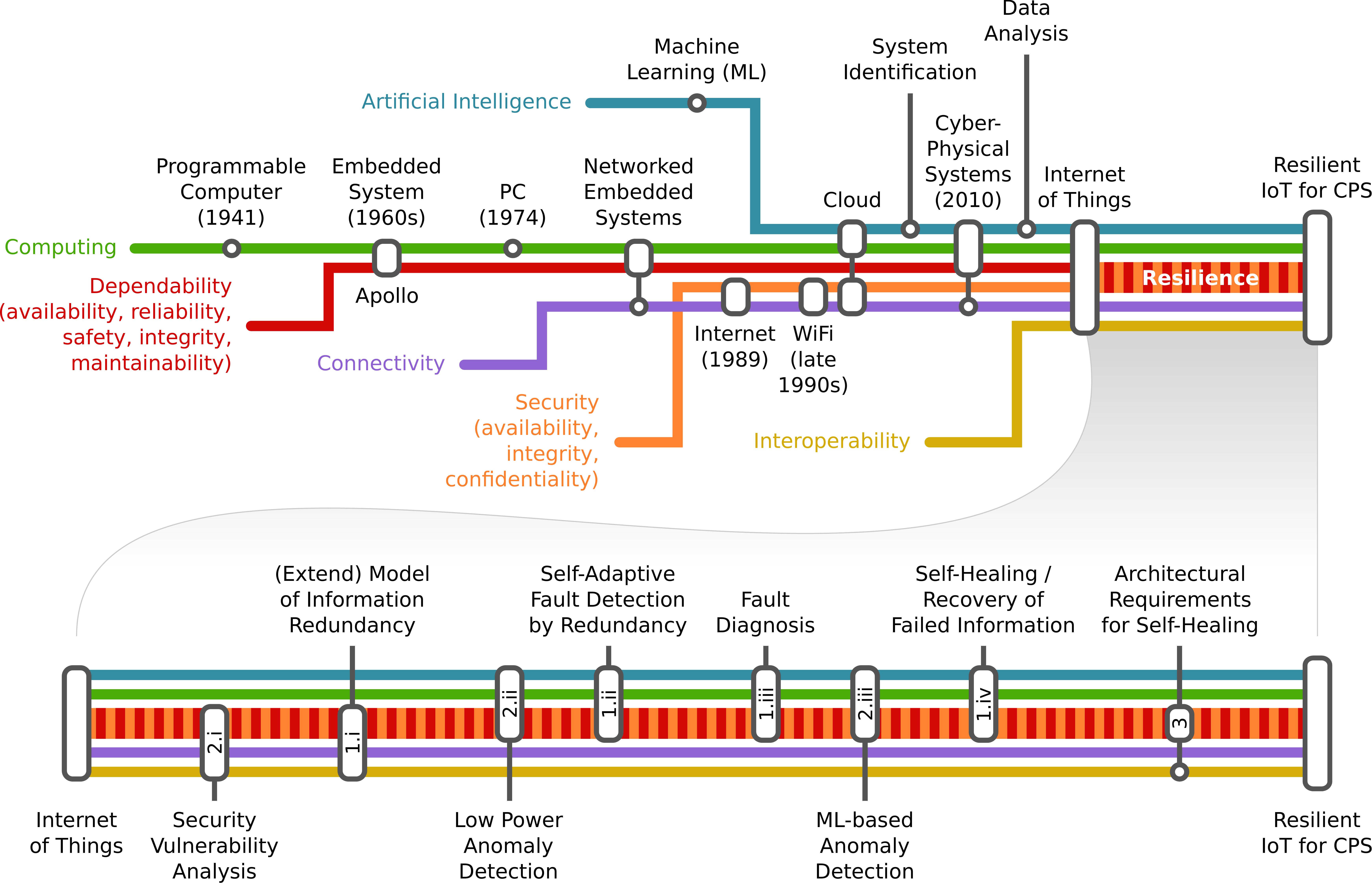}
  \caption{A brief history of computer systems and our roadmap towards resilient IoT for CPS.}
  \label{fig:roadmap}
\end{figure}

Figure~\ref{fig:roadmap} depicts the evolution of embedded systems (milestones as junctions),
their goals and requirements (as lines).
The lower part of Fig.~\ref{fig:roadmap} summarizes our milestones (1.i-iv, 2.i-iii, 3) given below.

\begin{itemize}
  \item[\textbf{1)}]
\textbf{How to improve the resilience of the IoT by fail-operational mechanisms?}
\textbf{How to monitor IoT components?}
\end{itemize}
The IoT will most likely contain many heterogeneous components with different capabilities of resilience.
We therefore consider fail-operational mechanisms that target
the \emph{dependability of the information exchanged} between IoT components.
The mechanism shall be applicable within the fog and/or cloud running on an independent component
or may be applied in an IoT device itself if the performance requirements
for the mechanism are satisfied.

We use given implicit redundancy of information provided by distinct IoT components to self-heal the IoT.
To this end, the major effort lies in
\emph{i)} developing and extending our redundancy model,
\emph{ii)} implementing a self-adaptive fault detection,
\emph{iii)} applying fault diagnosis, and
\emph{iv)} recovery considering currently available information.

\begin{itemize}
  \item[\textbf{2)}]
\textbf{How to verify IoT components?}
\textbf{How to detect anomalies in the IoT with minimum performance and energy overhead?}
\end{itemize}
We propose a methodology which consists of the following phases:
\emph{i)} security vulnerability analysis,
\emph{ii)} low-power and
\emph{iii)} ML-based anomaly detection.

The first phase of the proposed methodology is to analyze the IoT for CPS for the security vulnerabilities. Unlike the traditional simulation and emulation techniques, we plan to leverage the formal verification for analyzing the security vulnerabilities.
After identifying the security vulnerabilities and the corresponding parameters, i.e., communication and side-channel parameters, the next step is to use this information to develop online anomaly detection techniques. In this project, we plan to leverage two key characteristics, i.e., communication behavior and power (dynamic and leakage) to develop the low power and ML-based anomaly detection techniques.

\begin{itemize}
  \item[\textbf{3)}]
\textbf{What architectural requirements are necessary to ensure resilience with these mechanisms?}
\end{itemize}
Finally we collect the architectural requirements of our developed mechanisms
to be added to design guidelines for resilient IoT.

In the following, we present a case study to demonstrate
how the above mechanisms can be employed in a real-world use case
to detect, diagnose and mitigate faults.

\section{Case Study: Resilient Smart Mobility}
\label{sec:casestudy}

To illustrate the effectiveness of our approach,
we perform a case study on mobile autonomous systems, i.e.,
vehicle-to-everything~(V2X) communication in automated driving.
The network connects sensors, controllers and actuators, buildings, infrastructure and roadside systems.

\begin{figure}[htb]
  \centering
  \includegraphics[width=0.7\linewidth]{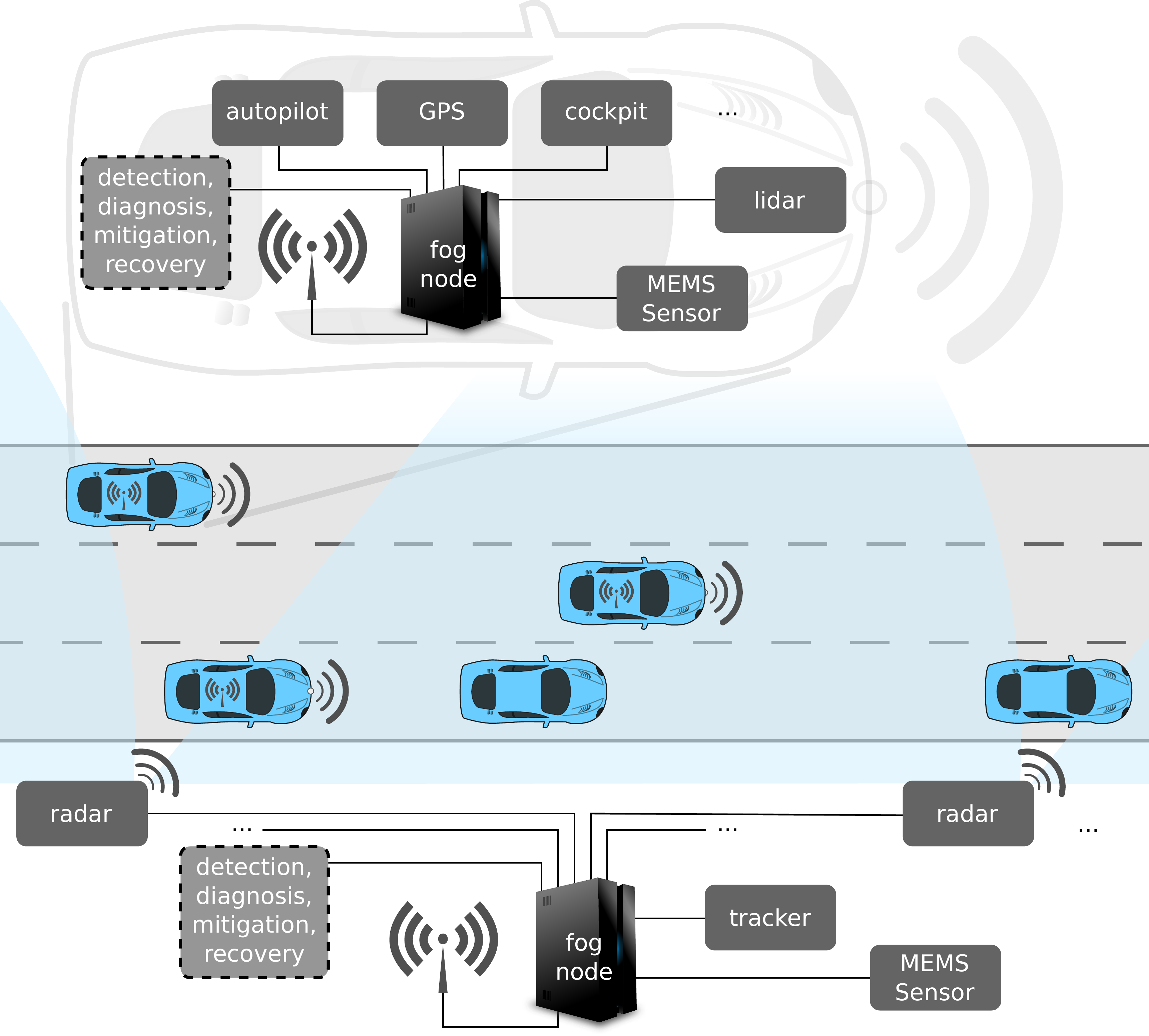}
  \caption{Visualization of the use case.}
  \label{fig:use-case}
\end{figure}

In particular, let's consider vehicles driving on a highway~(Fig.~\ref{fig:use-case}).
Radar sensors are mounted along the street and form a collaborative sensor field.
In order to improve object detection and classification,
a multi-object tracking scheme is employed,
which uses subsequent sensor measurements in the form of prediction and update cycles to estimate vehicle locations.
The tracking data can be used for, e.g., traffic congestion forecast or accident investigations.
A set of radar sensors is connected to a fog node, that is a computing unit and IoT gateway in
the near area of the sensors. The tracker - a software component running on a
fog node - tracks the vehicles on the road segment covered by the associated radars.
Some vehicles (e.g., autonomous cars) are equipped with distance
sensors like radar, lidar or depth cameras. The fog node(s) of these cars can
connect to near fog nodes of the street (directly over a vehicular network
called VANET, or via the mobile network over the cloud).
Additional MEMS sensors can support energy management, health and comfort in road transportation.

We assume the IoT infrastructure (things, fog, cloud, network) is given and
propose methods to increase the resilience of the IoT.

Failures of the radar sensors in our example will lead to inaccurate or even
unusable tracking results.
Failure scenarios like communication crashes and dead batteries (fail-silent,
fail-stop) are relatively easy to handle (e.g., watchdog/timeout). However, the
sensor measurements received by the tracker running in the fog node may be
erroneous due to noise (e.g., communication line, aging), environmental
influences (e.g., dirtying of the radar) or a security breach (e.g., hacked
fog node that collects data of a group of sensors). To detect a failure of
the sensor one has to create particular failure models for each possible hazard
(c.f., aging, dirtying and a security breach). A simple method detecting a
faulty sensor value in different failure scenarios is to check against other
information sources, i.e., exploit redundancy. However, explicit redundancy
that is replicating observation components is costly.

%

Self-healing can be applied to react also to failures not specifically
considered during design-time.
A very promising way of achieving self-healing is through structural
adaptation~(SHSA), by replacing a failed component with a substitute component
by exploiting implicit redundancy (or functional and temporal
redundancy)~\cite{Ratasich:2017}.
We use a knowledge base \cite{Hoeftberger:2015,Ratasich:2018} modeling
relationships among system variables given that certain implicit redundancy
exists in the system and extract a substitute from that knowledge base using
guided search~(Sec.~\ref{sec:roadmap-act}). The knowledge base can also be used
to monitor the system by comparing the information of variables against each
other, i.e., to detect failures~(Sec.~\ref{sec:roadmap-reason}).

\begin{figure}[htb]
  \centering
  \includegraphics[width=0.7\linewidth]{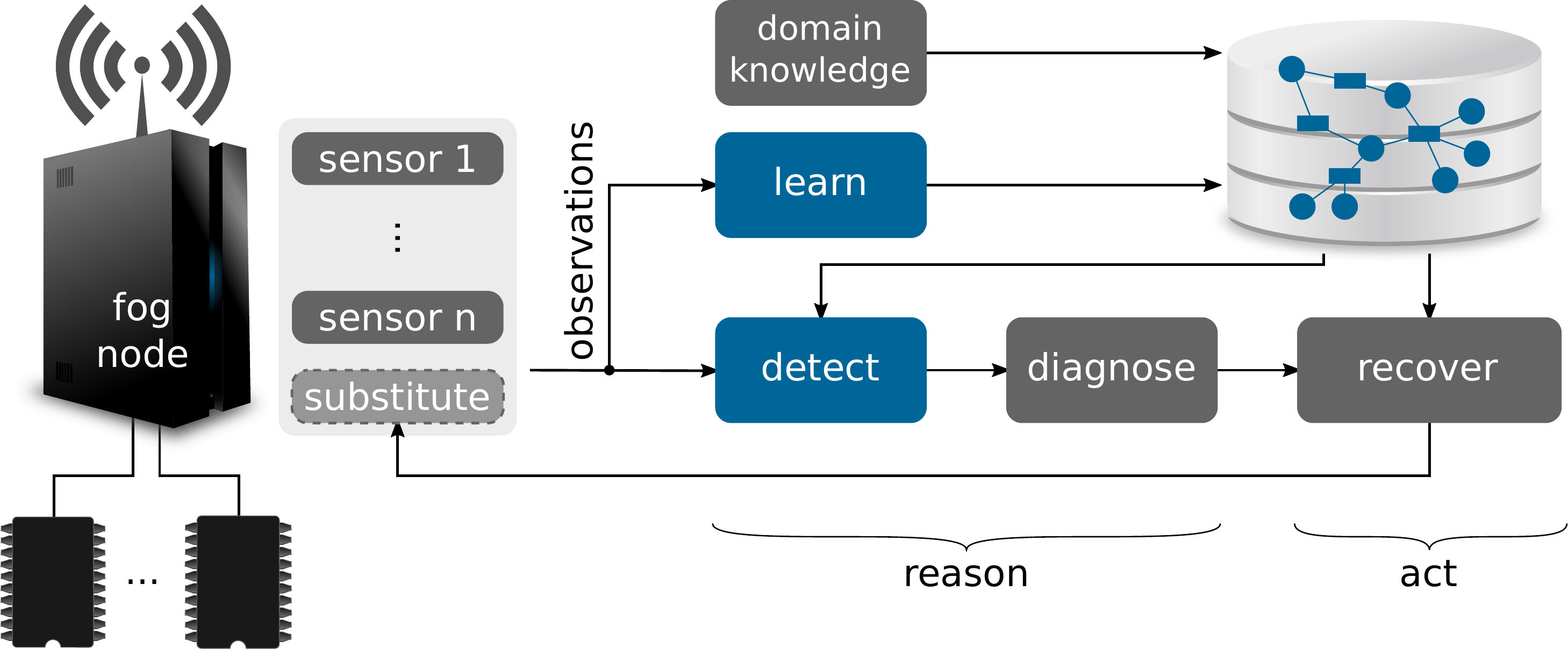}
  \caption{Overview of the self-healing components and proposed integration into a fog node.}
  \label{fig:self-healing}
\end{figure}

SHSA can be encapsulated in separate components listening and acting on the
communication network of the IoT, e.g., as tasks \emph{monitor},
\emph{diagnose} and \emph{recover} running on a fog
node~(Fig.~\ref{fig:self-healing}).

SHSA monitors the information communicated between components (typically the
sensor measurements or filtered/estimated observations), identifies the failed
component and replaces messages of the failed component delivering an erroneous
output by spawning a substitute software component.
SHSA considers the currently available information in the network, i.e., can be
applied in dynamic systems like the IoT (components may be added and removed
during runtime).
The knowledge base, in particular the relationships between the communicated
information, can be defined by the application's domain expert or learned
(approximated by, e.g., neural networks, SVMs or polynomial functions, see
also~\cite{Ratasich:2017}).

Alternatively, the monitor and diagnose task may be installed in the cloud
analyzing the logged tracks to trigger maintenance of radar sensors. The
requirements needed by SHSA regarding the architecture of the system (e.g.,
communication network) and a reference implementation of SHSA can be found
in~\cite{Ratasich:2017}.

\subsection{Detection and Diagnosis}
\label{sec:roadmap-reason}

In our future work, we want to use the SHSA knowledge base described
below to perform plausibility checks upon related
information.

%
%
As our focus is on adaptation of the software cyber-part in a CPS (cf. dynamic
reconfiguration of an FPGA), we assume that each physical component comprises
at least one software component (e.g., the driver of the radar in the vehicle)
and henceforth, consider the software components only.
The CPS implements certain functionality, e.g., a desired service (e.g., collision
avoidance). The subset of components implementing the CPS' objectives are
called controllers.

\subsubsection{SHSA Knowledge Base}

A system can be characterized by properties referred to as \emph{variables}
(e.g., the position and velocity of a tracked vehicle). The values of system
variables are communicated between different components typically via message-based
interfaces. Such transmitted data that is associated to a variable, we denote
as information atom, short \emph{itom}~\cite{Kopetz:2014}. A variable can be
provided by different components simultaneously (e.g., two radars with
overlapping field of view). Each software component executes a program that
uses input itoms and provides output itoms.
An itom is \emph{needed}, when it is input of a controller. A variable is
\emph{provided} when at least one corresponding itom can be received.

Variables are related to each other. A relation is a function or program (e.g.,
math, pseudo code or executable python code) to evaluate an output variable
from a set of input variables.

The knowledge base is a bipartite directed graph (which may also contain
cycles) with independent sets of variables and relations of a CPS. Variables
and relations are the nodes of the graph. Edges specify the input/output
interface of a relation.
For instance, Fig.~\ref{fig:knowledge_base} models the relationships between
the variables in the tracking use case (only relevant nodes, relationships and
edge directions for the scenario in Fig.~\ref{fig:scenario} are shown).
The knowledge base can also be encoded by a set of rules, e.g., written in Prolog.
It is then possible to further customize the model,
e.g., to follow the requirements and constraints of a CPS application.

A proper data association identifies which itoms or measurements represent the same variable,
e.g., links the different position itoms $(x, y, v)|_*$ to each other.
For instance, the GPS position $(x, y, v)|_{GPS}$ of a vehicle
(transmitted by the vehicle itself) has to be linked to the corresponding radar
track $(x, y, v)|_{radar}$ (provided by the radar).

Subsequently, the redundant itoms can be used, e.g., to monitor a radar sensor,
to substitute a failed radar or to increase the accuracy of a tracking
application by sensor fusion.
\begin{figure}[htb]
  \centering

\tikzstyle{var} = [draw, ellipse, inner sep=3pt, align=center]
\tikzstyle{rel} = [draw, rectangle, rounded corners, inner sep=3pt, minimum size = 1.8em]
\tikzstyle{itom} = [rectangle, align=center, font=\scriptsize]

\begin{tikzpicture}[auto, node distance=1em, >=latex, font=\sffamily\small]

  \node [var] (v1) {$(x, y, v)$};
  \node [itom, below = 0.5em of v1] (i1) {
    \textbf{GPS}\\
    \textbf{radar}\\
    $\mathrm{radar_{pre}}$%
  };

  \node [rel, left = of v1] (r1) {$\int$};
  \node [var, below left = of r1] (v21) {$(x', y', v')$};
  \node [itom, below = 0.5em of v21] (i2) {
    \textbf{GPS}\\
    \textbf{radar}\\
    $\mathbf{radar_{pre}}$%
  };
  \node [var, left = of r1] (v22) {$t'$};
  \node [var, above left = of r1] (v23) {$t$};

  \draw [->] (v21) -- (r1);
  \draw [->] (v22) -- (r1);
  \draw [->] (v23) -- (r1);
  \draw [->] (r1) -- (v1);

  \node [rel, right = of v1] (r2) {$+$};
  \node [var, above right = of r2] (v31) {$(x^{\triangleleft}, y^{\triangleleft}, v^{\triangleleft})$};
  \node [itom, right = 0.5em of v31] (i31) {
    \textbf{GPS}\\
    radar\\
    $\mathrm{radar_{pre}}$%
  };
  \node [var, below right = of r2] (v32) {$(d_x, d_y)$};
  \node [itom, below = 0.5em of v32] (i3) {
    \textbf{distance}%
  };

  \draw [->] (v31) -- (r2);
  \draw [->] (v32) -- (r2);
  \draw [->] (r2) -- (v1);

  \node [rel, right = of v32] (r4) {..};
  \node [var, right = of r4] (v4) {$D$};
  \node [itom, below = 0.5em of v4] (i4) {
    radar\\
    \textbf{lidar}\\
    camera system%
  };

  \draw [->] (v4) -- (r4);
  \draw [->] (r4) -- (v32);

\end{tikzpicture}
  \caption{Knowledge base. Ellipses are variables, boxes are relationships
    (functions). The variables are annotated with possible itoms. Bold itoms
    are available in the scenario in Fig.~\ref{fig:scenario}.}
  \label{fig:knowledge_base}
\end{figure}
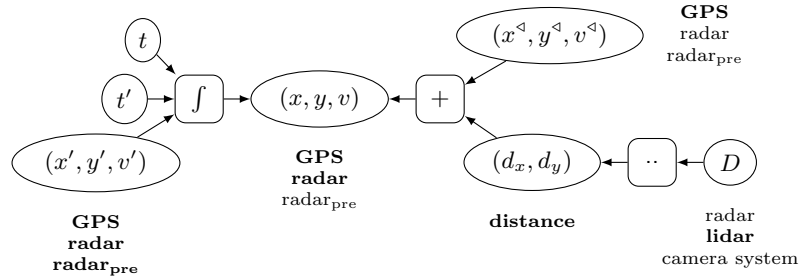
\begin{figure}[htb]
  \centering
  \input{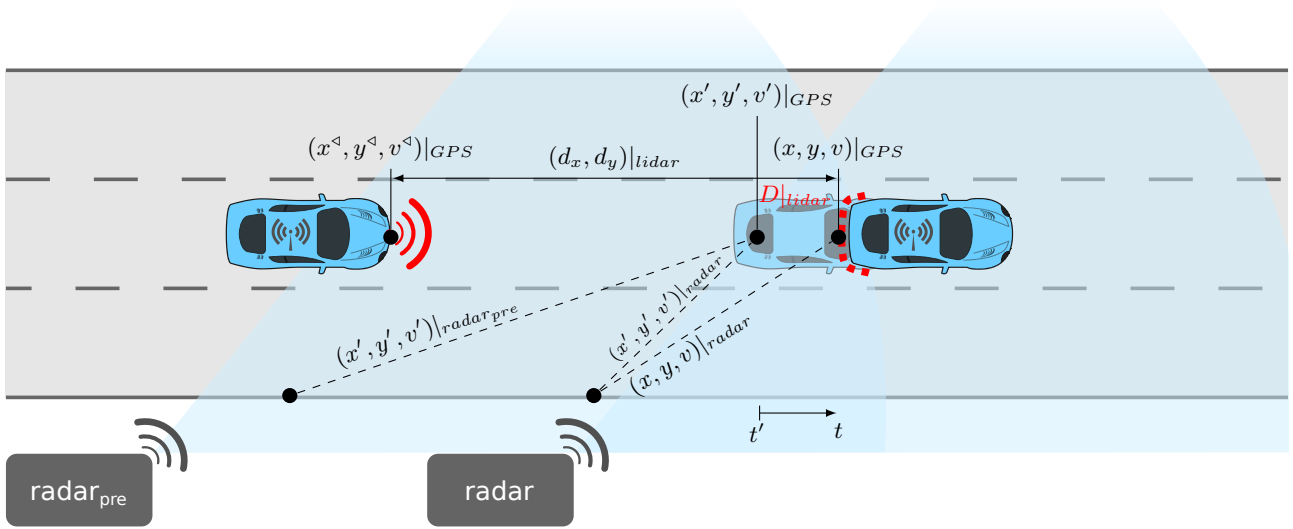}
  \caption{An exemplary scenario from the use case.
    Visualization of itoms~$variable|_{itom}$ from the knowledge base in Fig.~\ref{fig:knowledge_base}.}
  \label{fig:scenario}
\end{figure}

The interested reader is referred to \cite{Ratasich:2018} and \cite{Hoeftberger:2015}
for more details on the SHSA knowledge base.

\subsubsection{Fault Detection by Redundancy}
An itom has failed, when it deviates from the specification.
Our monitor uses the knowledge base to periodically perform a plausibility check
to identify a failed itom.
The automatic setup of a runtime monitor follows successive procedure:
\begin{itemize}
  \item Select the variable to be monitored (typically the corresponding
    variable to the itom under test), e.g., the position of a vehicle.
  \item Collect the provided itoms (e.g., subscribe to all available
    messages). Note, the availability of variables may change from time to time which
    should trigger a new setup of the monitor.
  \item Extract relations of the monitored variable and available variables
    from the knowledge base (similar to the search of valid substitutions in Sec.~\ref{sec:roadmap-act}).
\end{itemize}

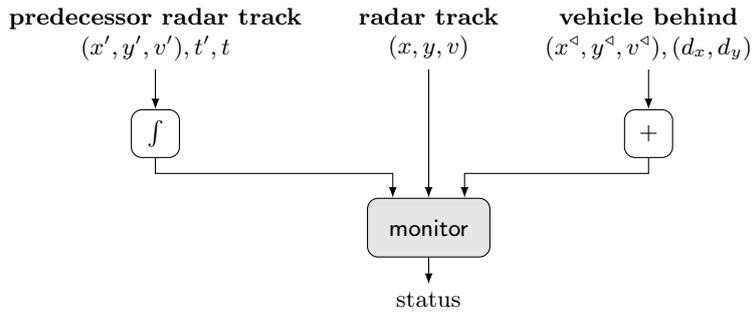
\begin{figure}[htb]
  \centering

\tikzstyle{component} = [draw, fill=black!10, rectangle, rounded corners, inner sep=8pt]
\tikzstyle{io} = [rectangle, inner sep=3pt, align=center, font=\small]

\tikzstyle{var} = [draw, ellipse, inner sep=3pt, align=center]
\tikzstyle{rel} = [draw, rectangle, rounded corners, inner sep=3pt, minimum size = 1.8em]
\tikzstyle{itom} = [rectangle, align=center, font=\small]

\begin{tikzpicture}[auto, node distance=1em, >=latex, font=\sffamily\small]
  \node [component] (monitor) {monitor};
  \node [io, below = of monitor] (status) {status};
  \draw [->] (monitor) -- (status);

  \node [itom, above = 4.85em of monitor] (itest) {
    \textbf{radar track}\\
    $(x, y, v)$%
  };
  \draw [->] (itest) -- (monitor);

  \node [rel, above left = 1.5em and 7em of monitor] (r1) {$\int$};
  \node [itom, above = 1.5em of r1] (i1) {
    \textbf{predecessor radar track}\\
    $(x', y', v'), t', t$%
  };
  \draw [->] (i1) -- (r1);
  \draw [->] (r1) |- +(0,-1.5em) -| (monitor.140);

  \node [rel, above right = 1.5em and 5em of monitor] (r2) {$+$};
  \node [itom, above = 1.5em of r2] (i21) {
    \textbf{vehicle behind}\\
    $(x^{\triangleleft}, y^{\triangleleft}, v^{\triangleleft}), (d_x, d_y)$%
  };
  \draw [->] (i21) -- (r2);
  \draw [->] (r2) |- +(0,-1.5em) -| (monitor.40);
\end{tikzpicture}
  \caption{A monitor checking the position of a vehicle using different itoms. The
    itoms are first transferred into the common domain (here: position of the
    vehicle $(x, y, v)$) and compared against each other.}
  \label{fig:monitor}
\end{figure}
The instantiated monitor for the position of a vehicle is depicted in
Fig.~\ref{fig:monitor}.
At each time step the relations are executed to bring the available itoms
(provided variables) into the common domain (variable to be monitored) where the
values are compared against each other.
The monitor returns the fault status or a confidence / health / trust value for
each itom used in the plausibility check.

The confidence may be expressed by a distance metric or error between the itoms
in the common domain.
The trust or confidence of a radar may be accumulated from the individual
confidence values of the tracked vehicles, i.e., the vehicles in the field of
view of the radar. As soon as the confidence falls below a specific threshold
for a specific amount of time the status of the respective itom is classified as failed.

The monitor can identify failed itoms in the common domain, however, when the
output of a relation mismatches in the common domain, all inputs of the
relation are marked faulty. To avoid additional monitors (a monitor for each
input variable is necessary to identify the failed itom)
a fault localization can be performed.

\subsubsection{Anomaly Detection}

In addition to SHSA, we are developing a low power runtime anomaly detection
and ML-based runtime anomaly detection~\cite{shafique2018intelligent} to ensure:
\emph{i)} a secure and safe platform for automated driving, and
\emph{ii)} secure V2X communication.

\paragraph{Low Power}
To address the key challenge of power overhead in CPS,
we propose a methodology that leverages the traditional low-power online anomaly detection techniques,
in particular, assertions, sensor-based analysis and runtime monitoring.
In this methodology, the first step is to identify an appropriate detection scheme based on the security threats, security metrics and design constraints.
Second, based on the selected technique, the setup of corresponding assertions or sensor-based runtime monitoring is developed and implemented.

\begin{figure}[htb]
  \centering
  \includegraphics[width=0.99\linewidth]{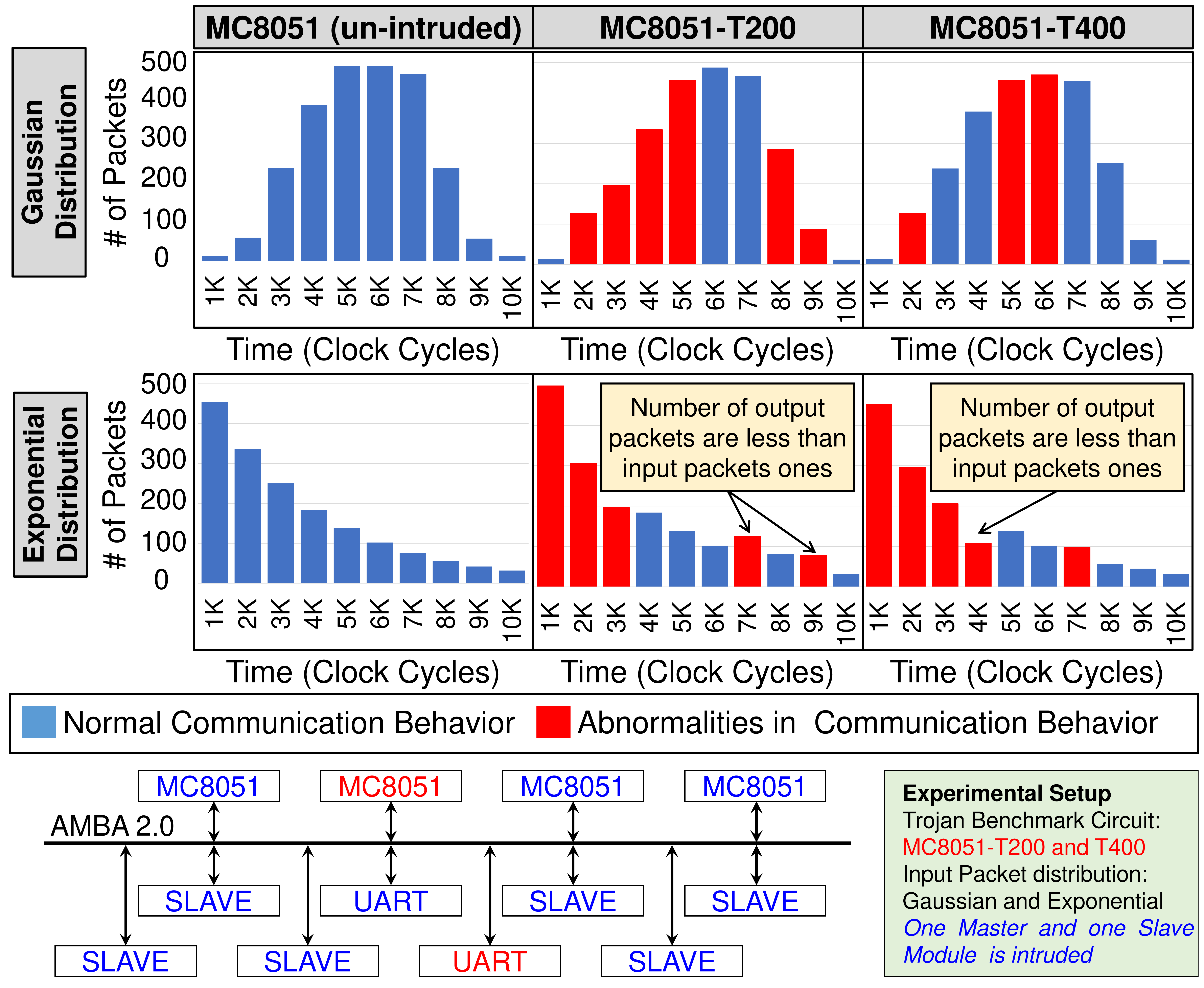}
  \caption{The effects of trust-hub Trojan benchmarks (i.e., MC8051-T200 and T400) on the communication behavior of MC8051 for Gaussian and Exponential input data distribution and an Overview of the motivational case study of an MC8051-based communication network.}
  \label{fig:Com_analysis}
\end{figure}

We propose to use communication behavior-based assertions to identify the online anomalies with low power and area overhead.
To illustrate the effect of intrusions on communication behavior, we analyzed the effects of several MC8051 trust-Hub benchmarks on an MC8051-based communication network.
The analysis in Fig. \ref{fig:Com_analysis} shows that in case of a denial-of-service attack, output packets of the communication channel are less than the input ones.
However, in case of flooding, jamming, and information leakage attacks the traffic in the communication channel is more than the input data injection. Therefore, it can be concluded that the communication behavior can be used to identify the anomalous behavior.
However, extracting the communication behavior without any golden circuit is not straight-forward which raises the following research challenges:

\begin{enumerate}
\item How to identify the reference communication behavior without any reference circuits/systems?
\item How to statistically model the communication behavior which can be used to identify the anomalous behavior?
\item How to measure and analyze the communication for low power runtime anomaly detection?
\end{enumerate}

\paragraph{Machine Learning}
With the increasing trend of connected devices, the number of communication channels also increases exponentially. Thus, communication behavior-based assertions are not feasible to handle the large number of communication channels and corresponding communication data. Therefore, to increase the scope of the online anomaly detection for larger CPS with big data analysis, in this project, we propose to explore machine learning algorithms to extract the hidden features from the side-channel parametric and communication behaviors (i.e., power and communication behavior). The first step to develop a ML-based anomaly detection is to select an appropriate ML algorithm based on the design constraints, security threats and complexity of the measured data. Then, train and implement the ML algorithm based on measured data with minimum power and area overhead. 

\begin{figure}[htb]
  \centering
  \includegraphics[width=0.99\linewidth]{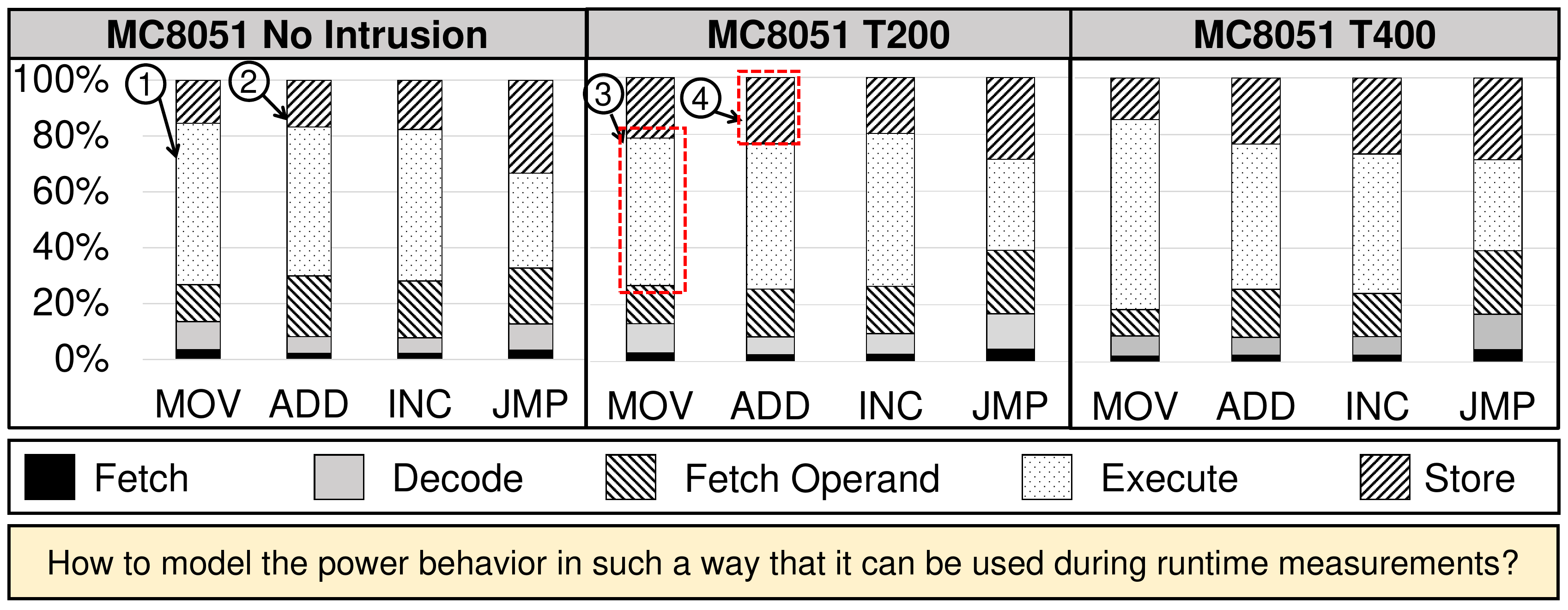}
  \caption{Effects of trust-hub Trojan benchmarks (i.e., MC8051-T200, T300 and T400) on Power Correlation with respect to Pipeline Stages for Different Instructions, i.e., MOV, ADD, INC, JMP.}
  \label{fig:Power_analysis}
\end{figure}

To illustrate the effect of intrusions on power profile,
we analyzed the MC8051 \emph{with} and \emph{without} trust-Hub benchmarks,
i.e., MC8051-T200 and T400, in Xilinx power analyzer.
The experimental analysis in Fig.~\ref{fig:Power_analysis} shows that
intrusions in MC8051 have a significant impact on the power distribution
with respect to different pipeline stages (see labels 1 to 4).
Therefore, it can be concluded that the power profiling of the processing elements/controllers in CPS can be used to identify the abnormalities.

Though the power profiling of the microprocessor can be used to detect an anomalous behavior,
power-based ML training and runtime measurement is not easy.
Therefore, the following research challenges must be considered while designing the ML-based online anomaly detection:
\begin{enumerate}
\item How to extract the power profiles of the processing elements (controllers in CPS) for efficient ML training? 
\item How to reduce the area and energy overhead of power-ports for runtime measurement and modeling?
\end{enumerate}

\subsubsection{Fault Localization~\texorpdfstring{\cite{BartocciFMN18}}{}}
The fault detection mechanisms described in the last sections can identify failed data on the communication network.
In order to recover the failed component responsible for the wrong information we have to apply fault localization.

The engineers often design CPS using the MathWorks$\texttrademark$ Simulink toolset 
to model their functionalities.  These models are generally complex hybrid systems 
that are often impossible to analyze only by using the reachability analysis techniques 
described before.  A popular technique to find bugs in Simulink/Stateflow models is 
falsification-based testing~\cite{staliro,breach,0001SDKJ15}.
This approach consists in monitoring
an STL property over traces produced by systematically simulating the CPS 
design using different set of test cases.  For each generated trace the monitor
returns a real-value that provides an indication as how far the trace is from 
violation.  This information can be used to guide the test case generation 
to find an input sequence that would falsify the specification.  
However, this approach does not provide any information concerning 
which is the failed component and the precise moment in time that is 
responsible for the observed violation.  To overcome this shortcoming, 
in~\cite{BartocciFMN18} Bartocci \etal{} have recently introduced a new 
procedure that aids designers in debugging Simulink/Stateflow hybrid 
system models, guided by STL specifications.  This approach 
combines a trace diagnostics~\cite{FerrereMalerNickovic15}  technique that localizes  time 
segments and interface variables contributing to the 
property violations, a slicing method~\cite{ReichertG12}  that maps these time segments 
to the internal states and transitions of the model and  a 
spectrum-based fault-localization method~\cite{4344104} that produces 
a ranking of the internal states and/or transitions that are
most likely to explain the fault.

\subsection{Recovery or Mitigation}
\label{sec:roadmap-act}

A failed itom can be replaced by a function of related itoms. To this end, the
knowledge base is searched for relationships using provided variables and
spawns a substitute.

\subsubsection{Replacement}

The substitute search algorithm traverses the knowledge base
(Fig.~\ref{fig:knowledge_base}) from the failed but needed information as root
to find a valid substitution~\cite{Ratasich:2018}.

%
%
A substitution of a variable is a connected acyclic sub-graph of the knowledge
base with following properties: i) The output variable is the only sink of the
substitution. ii) Each variable has zero or one relationship as
predecessor. iii) All input variables of a relation must be included (it
follows that the sources of the substitution graph are variables only).

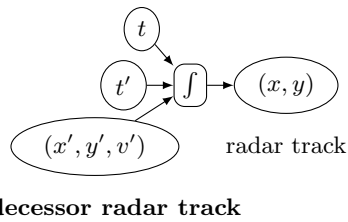
\begin{figure}[htb]
  \centering

\tikzstyle{var} = [draw, ellipse, inner sep=3pt, align=center]
\tikzstyle{rel} = [draw, rectangle, rounded corners, inner sep=3pt]
\tikzstyle{itom} = [rectangle, align=center, font=\small]
\tikzstyle{active} = [draw=blue]
\tikzstyle{failed} = [draw=red]

\begin{tikzpicture}[auto, node distance=1em, >=latex, font=\sffamily\small]
  \node [var] (v1) {$(x, y)$};
  \node [itom, below = 0.5em of v1] (i1) {
    radar track%
  };

  \node [rel, left = of v1] (r1) {$\int$};
  \node [var, above left = of r1] (v23) {$t$};
  \node [var, left = of r1] (v21) {$t'$};
  \node [var, below left = of r1] (v22) {$(x', y', v')$};
  \node [itom, below = 0.5em of v22] (i2) {
    \textbf{predecessor radar track}%
  };

  \draw [->] (v21) -- (r1);
  \draw [->] (v22) -- (r1);
  \draw [->] (v23) -- (r1);
  \draw [->] (r1) -- (v1);
\end{tikzpicture}
  \caption{A valid substitution for the failed street radar.
    Old data from the predecessor radar is used to forward estimate the position of the vehicles.}
  \label{fig:substitution}
\end{figure}

A substitution is valid if all sources are provided, otherwise the substitution
is invalid~(Fig.~\ref{fig:substitution}). Only a valid substitution can be
instantiated (to a substitute) by concatenating the relationships which take
the selected itoms as input (e.g., best itoms of the source variables).

Substitutions can be found by depth-first search of the knowledge base with the
failed variable as root. The search may stop as soon as all unprovided
variables are substituted~\cite{Hoeftberger:2015}. In~\cite{Ratasich:2018} we
present a guided search approach using a performance measure for substitutions.


The result of the search - the substitution - is instantiated in a substitute~\cite{Ratasich:2017}.
In particular, the substitute subscribes to the input itoms and
concatenates the functions or programs from the relationships. The substitute
then periodically publishes the output.
To avoid inconsistencies and fault propagation, the failed component (probably
publishing erratic messages) should be shut down as soon as possible.

\section{Conclusion}
\label{sec:conclusion}

This paper summarizes the state-of-the-art
of detection and recovery to react to failures in IoT for CPS.
We further presented the main challenges and a roadmap towards a resilient IoT.
The summary of the main challenges identified for existing and new resilience methods are:
\begin{itemize}
\item Limited resources of computation and power (e.g., for runtime data acquisition).
\item Limited knowledge of device and interface semantics
  (e.g., to retrieve a reference behavior for anomaly detection
  or model the redundancies in the system).
\item Ensure and do not alter (real-time) behavior
  by adding or applying resilience techniques.
\item Provide long-term dependability and security, that is,
  ensure resilience also after environmental, functional or technological changes of the system.
\item Adaptation, verification, validation and robustness of the resilience techniques.
\end{itemize}

Moreover, we introduced some of our key solutions on an automotive example.
The SHSA knowledge base presented in Section~\ref{sec:casestudy}
describes implicit and explicit redundancy in a
communication network. It can therefore be exploited to monitor, replace or
fuse information.
%
Because SHSA is based on redundancy it can handle various fault scenarios.
Especially permanent faults in the IoT can be detected and recovered
given some redundancy exists.
As long as the failed components can be isolated and replaced by redundant information
the methods can handle physical, development or interaction faults
manifested as failures at the components' interfaces.

The monitors tackle the requirement on fault detection by voting over redundant information or comparing it to some reference behavior~(R1).
An additional fault localization identifies and triggers a disconnection of the failed component to avoid fault propagation.
The substitution replaces failed information with redundant one~(R2).

The presented techniques need a reference behavior,
common understanding of the information or access to relevant redundancy (C2).
Therefore, the IoT should provide proper interoperability (e.g., in form of standards).
Under some constraints (bounded or static SHSA knowledge base, estimation of the worst-case execution time of relationships) SHSA is suitable for real-time applications~\cite{Hoeftberger:2015}.
However, solutions to increase scalability have to be investigated (C3).
Moreover, the individual IoT devices might not have the resources to implement detection and recovery (C1).
In future work we therefore want to focus on a distributed approach of the mechanism
(e.g., by splitting the knowledge base for subsystems,
or monitor in a distributed fashion like agreement protocols do).

\section*{Acknowledgement}
The research leading to these results has received funding from the IoT4CPS
project partially funded by the ``ICT of the Future'' Program of the FFG and
the BMVIT.
The authors acknowledge the TU Wien University Library for financial support through its Open Access Funding Programme.

\bibliographystyle{unsrt}
\bibliography{bib/denise,bib/ezio,bib/faiq}

\begin{thebibliography}{100}

\bibitem{Lee:2010}
Edward~A. Lee and Sanjit~A. Seshia.
\newblock An introductory textbook on cyber-physical systems.
\newblock In {\em Proceedings of the 2010 Workshop on Embedded Systems
  Education}, WESE '10, pages 1:1--1:6, New York, NY, USA, 2010. ACM.

\bibitem{Ragunathan2010}
Ragunathan~(Raj) Rajkumar, Insup Lee, Lui Sha, and John Stankovic.
\newblock Cyber-physical systems: The next computing revolution.
\newblock In {\em Proc. of DAC '10: the 47th Design Automation Conference},
  pages 731--736, New York, NY, USA, 2010. ACM.

\bibitem{Rajkumar2012}
Ragunathan Rajkumar.
\newblock A cyber-physical future.
\newblock {\em Proceedings of the IEEE}, 100(Special Centennial
  Issue):1309--1312, 2012.

\bibitem{Ceccarelli:2016}
Andrea Ceccarelli, Andrea Bondavalli, Bernhard Froemel, Oliver Hoeftberger, and
  Hermann Kopetz.
\newblock {\em Basic Concepts on Systems of Systems}, pages 1--39.
\newblock Springer International Publishing, Cham, 2016.

\bibitem{Fagnant2015}
Daniel~J. Fagnant and Kara Kockelman.
\newblock Preparing a nation for autonomous vehicles: opportunities, barriers
  and policy recommendations.
\newblock {\em Transportation Research Part A: Policy and Practice},
  77:167--181, 2015.

\bibitem{Sokolsky2012}
Insup Lee, Oleg Sokolsky, Sanjian Chen, John Hatcliff, Eunkyoung Jee, BaekGyu
  Kim, Andrew King, Margaret Mullen-Fortino, Soojin Park, Alexander Roederer,
  and Krishna~K. Venkatasubramanian.
\newblock Challenges and research directions in medical cyber-physical systems.
\newblock {\em Proceedings of the IEEE}, 100(1):75--90, 2012.

\bibitem{statistica2017iot}
IHS.
\newblock {Internet of Things (IoT) connected devices installed base worldwide
  from 2015 to 2025 (in billions)}.
\newblock In {\em Statista - The Statistics Portal}.
\newblock Accessed September 10, 2018. Available from
  \url{https://www.statista.com/statistics/471264/iot-number-of-connected-devices-worldwide/}.

\bibitem{statistica2018population}
{World Bank}.
\newblock {World: Total population 2007-2017 (in billion inhabitants)}.
\newblock In {\em Statista - The Statistics Portal}.
\newblock Accessed October 18, 2018. Available from
  \url{https://www.statista.com/statistics/805044/total-population-worldwide/}.

\bibitem{reinsel2017data}
David Reinsel, John Gantz, and Johnl Rydning.
\newblock Data age 2025: The evolution of data to life-critical don't focus on
  big data; focus on data that's big.
\newblock In {\em IDC, Seagate, April}.
\newblock Accessed October 19, 2018. Available from
  \url{https://www.seagate.com/files/www-content/our-story/trends/files/Seagate-WP-DataAge2025-March-2017.pdf}.

\bibitem{intel2018cars}
Kathy Winter.
\newblock For self-driving cars, there's big meaning behind one big number: 4
  terabytes.
\newblock In {\em Intel News Room}.
\newblock Accessed October 19, 2018. Available from
  \url{https://newsroom.intel.com/editorials/self-driving-cars-big-meaning-behind-one-number-4-terabytes/
  }.

\bibitem{nytimes2015}
Jad Mouawad.
\newblock {F.A.A. Orders Fix for Possible Power Loss in Boeing 787}.
\newblock
  \href{https://www.nytimes.com/2015/05/01/business/faa-orders-fix-for-possible-power-loss-in-boeing-787.html}{New
  York Times, May 1}, 2015.

\bibitem{washingtonpost2018}
Peter Holley.
\newblock {Chrysler Fiat announces recall of nearly 5 million U.S. cars}.
\newblock
  \href{https://www.washingtonpost.com/news/innovations/wp/2018/05/25/chrysler-fiat-announces-recall-for-nearly-5-million-u-s-cars/?noredirect=on&utm_term=.ecb1a2e4d38a}{The
  Guardian, May 25}, 2018.

\bibitem{guardian2018}
Edward Helmore.
\newblock {Uber shuts down self-driving operation in Arizona after fatal
  crash}.
\newblock
  \href{https://www.theguardian.com/technology/2018/may/23/uber-shuts-down-self-driving-operation-in-arizona-two-months-after-fatal-crash}{The
  Guardian, May 23}, 2018.

\bibitem{Bloomfield:2013}
Robin Bloomfield, Kateryna Netkachova, and Robert Stroud.
\newblock Security-informed safety: If it's not secure, it's not safe.
\newblock In Anatoliy Gorbenko, Alexander Romanovsky, and Vyacheslav
  Kharchenko, editors, {\em Software Engineering for Resilient Systems}, pages
  17--32, Berlin, Heidelberg, 2013. Springer Berlin Heidelberg.

\bibitem{guardian2017}
Alex Hern.
\newblock {Hacking risk leads to recall of 500,000 pacemakers due to patient
  death fears }.
\newblock
  \href{https://www.theguardian.com/technology/2017/aug/31/hacking-risk-recall-pacemakers-patient-death-fears-fda-firmware-update}{The
  Guardian, August 31}, 2017.

\bibitem{bbc2016}
Rory Cellan-Jones.
\newblock {F.A.A. Orders Fix for Possible Power Loss in Boeing 787}.
\newblock \href{https://www.bbc.co.uk/news/technology-35667989}{BBC, February
  26}, 2016.

\bibitem{wired2015}
Andy Greenberg.
\newblock {Hackers Remotely Kill a Jeep on the Highway -- With Me in It}.
\newblock
  \href{https://www.wired.com/2015/07/hackers-remotely-kill-jeep-highway/}{Wired,
  July 1}, 2015.

\bibitem{Kolias2017}
Constantinos Kolias, Georgios Kambourakis, Angelos Stavrou, and Jeffrey Voas.
\newblock {D}{D}o{S} in the {I}o{T}: {M}irai and other {B}otnets.
\newblock {\em Computer}, 50(7):80--84, 2017.

\bibitem{Laprie:2008}
Jean-Claude Laprie.
\newblock {From Dependability to Resilience}.
\newblock In {\em Dependable Systems and Networks (DSN 2008), 38th Annual
  IEEE/IFIP International Conference}, 2008.

\bibitem{Atzori:2010}
Luigi Atzori, Antonio Iera, and Giacomo Morabito.
\newblock {The Internet of Things: A survey}.
\newblock {\em Computer Networks}, 54(15):2787 -- 2805, 2010.

\bibitem{Vermesan:2011}
Ovidiu Vermesan, Peter Friess, Patrick Guillemin, Sergio Gusmeroli, Harald
  Sundmaeker, Alessandro Bassi, Ignacio~Soler Jubert, Margaretha Mazura, Mark
  Harrison, Markus Eisenhauer, et~al.
\newblock Internet of things strategic research roadmap.
\newblock {\em Internet of Things-Global Technological and Societal Trends},
  1(2011):9--52, 2011.

\bibitem{wollschlaeger2017future}
Martin Wollschlaeger, Thilo Sauter, and Juergen Jasperneite.
\newblock The future of industrial communication: Automation networks in the
  era of the internet of things and industry 4.0.
\newblock {\em IEEE Industrial Electronics Magazine}, 11(1):17--27, 2017.

\bibitem{butun2014survey}
Ismail Butun, Salvatore~D Morgera, and Ravi Sankar.
\newblock A survey of intrusion detection systems in wireless sensor networks.
\newblock {\em IEEE communications surveys \& tutorials}, 16(1):266--282, 2014.

\bibitem{mitchell2014survey}
Robert Mitchell and Ing-Ray Chen.
\newblock A survey of intrusion detection techniques for cyber-physical
  systems.
\newblock {\em ACM Computing Surveys (CSUR)}, 46(4):55, 2014.

\bibitem{Buczak:2016}
A.~L. Buczak and E.~Guven.
\newblock {A Survey of Data Mining and Machine Learning Methods for Cyber
  Security Intrusion Detection}.
\newblock {\em IEEE Communications Surveys Tutorials}, 18(2):1153--1176,
  Secondquarter 2016.

\bibitem{Wu:2010}
Shelly~Xiaonan Wu and Wolfgang Banzhaf.
\newblock The use of computational intelligence in intrusion detection systems:
  A review.
\newblock {\em Applied Soft Computing}, 10(1):1 -- 35, 2010.

\bibitem{Avizienis:2004}
Algirdas Avizienis, Jean-Claude Laprie, Brian Randell, and Carl Landwehr.
\newblock {Basic Concepts and Taxonomy of Dependable and Secure Computing}.
\newblock {\em IEEE Trans. on Dependable and Secure Computing}, 1:11--33, 2004.

\bibitem{Kopetz:2011}
Hermann Kopetz.
\newblock {\em {Real-Time Systems: Design Principles for Distributed Embedded
  Applications}}.
\newblock Springer, New York, 2nd edition, 2011.

\bibitem{lncs10457}
Ezio Bartocci and Yli{\`{e}}s Falcone, editors.
\newblock {\em Lectures on Runtime Verification - Introductory and Advanced
  Topics}, volume 10457 of {\em Lecture Notes in Computer Science}.
\newblock Springer, 2018.

\bibitem{Leucker:2009}
Martin Leucker and Christian Schallhart.
\newblock A brief account of runtime verification.
\newblock {\em The Journal of Logic and Algebraic Programming}, 78(5):293 --
  303, 2009.

\bibitem{Chandola:2009}
Varun Chandola, Arindam Banerjee, and Vipin Kumar.
\newblock {Anomaly Detection: A Survey}.
\newblock {\em ACM Comput. Surv.}, 41(3):15:1--15:58, July 2009.

\bibitem{Cheng:2009}
{B. H. C. Cheng \emph{et al.}}
\newblock {Software Engineering for Self-Adaptive Systems: A Research Roadmap}.
\newblock In {\em Software Engineering for Self-Adaptive Systems}, pages 1--26.
  Springer Verlag, Berlin, Heidelberg, 2009.

\bibitem{Siva:2010}
S.~Siva Sathya and K.~Syam Babu.
\newblock Survey of fault tolerant techniques for grid.
\newblock {\em Computer Science Review}, 4(2):101 -- 120, 2010.

\bibitem{deLemos:2013}
{R. de Lemos \emph{et al.}}
\newblock {\em {Software Engineering for Self-Adaptive Systems: A Second
  Research Roadmap}}, pages 1--32.
\newblock Springer Berlin Heidelberg, Berlin, Heidelberg, 2013.

\bibitem{Weyns:2017}
Danny Weyns.
\newblock {Software Engineering of Self-Adaptive Systems: An Organised Tour and
  Future Challenges}.
\newblock Springer, 2017.

\bibitem{Kounev:2017}
Samuel Kounev, Peter Lewis, Kirstie~L. Bellman, Nelly Bencomo, Javier Camara,
  Ada Diaconescu, Lukas Esterle, Kurt Geihs, Holger Giese, Sebastian G{\"o}tz,
  Paola Inverardi, Jeffrey~O. Kephart, and Andrea Zisman.
\newblock {\em {The Notion of Self-aware Computing}}, pages 3--16.
\newblock Springer International Publishing, Cham, 2017.

\bibitem{Papp:2016}
Zoltan Papp and George Exarchakos, editors.
\newblock {\em {Runtime Reconfiguration in Networked Embedded Systems - Design
  and Testing Practices}}.
\newblock Internet of Things - Technology, Communications and Computing.
  Springer Science+Business Media Singapore, 2016.

\bibitem{humayed2017cyber}
Abdulmalik Humayed, Jingqiang Lin, Fengjun Li, and Bo~Luo.
\newblock Cyber-physical systems security -- a survey.
\newblock {\em IEEE Internet of Things Journal}, 4(6):1802--1831, 2017.

\bibitem{Isermann:2006}
Rolf Isermann.
\newblock {\em Fault-diagnosis systems: an introduction from fault detection to
  fault tolerance}.
\newblock Springer Science \& Business Media, 2006.

\bibitem{Ghosh:2007}
Debanjan Ghosh, Raj Sharman, H.~Raghav~Rao, and Shambhu Upadhyaya.
\newblock {Self-healing Systems - Survey and Synthesis}.
\newblock {\em Decis. Support Syst.}, 42(4):2164--2185, January 2007.

\bibitem{Psaier:2011}
Harald Psaier and Schahram Dustdar.
\newblock A survey on self-healing systems: approaches and systems.
\newblock {\em Computing}, 91(1):43--73, Jan 2011.

\bibitem{Elhady:2018}
Nancy~E ElHady and Julien Provost.
\newblock A systematic survey on sensor failure detection and fault-tolerance
  in ambient assisted living.
\newblock {\em Sensors}, 18(7):1991, 2018.

\bibitem{Zanella:2014}
Andrea Zanella, Nicola Bui, Angelo Castellani, Lorenzo Vangelista, and Michele
  Zorzi.
\newblock Internet of things for smart cities.
\newblock {\em IEEE Internet of Things journal}, 1(1):22--32, 2014.

\bibitem{Conoscenti:2016}
Marco Conoscenti, Antonio Vetro, and Juan~Carlos De~Martin.
\newblock Blockchain for the internet of things: A systematic literature
  review.
\newblock In {\em Computer Systems and Applications (AICCSA), 2016 IEEE/ACS
  13th International Conference of}, pages 1--6. IEEE, 2016.

\bibitem{ray2018survey}
Partha~Pratim Ray.
\newblock A survey on internet of things architectures.
\newblock {\em Journal of King Saud University-Computer and Information
  Sciences}, 30(3):291--319, 2018.

\bibitem{reyna2018blockchain}
Ana Reyna, Cristian Mart{\'\i}n, Jaime Chen, Enrique Soler, and Manuel
  D{\'\i}az.
\newblock On blockchain and its integration with iot. challenges and
  opportunities.
\newblock {\em Future Generation Computer Systems}, 2018.

\bibitem{khan2018iot}
Minhaj~Ahmad Khan and Khaled Salah.
\newblock Iot security: Review, blockchain solutions, and open challenges.
\newblock {\em Future Generation Computer Systems}, 82:395--411, 2018.

\bibitem{sfar2018roadmap}
Arbia~Riahi Sfar, Enrico Natalizio, Yacine Challal, and Zied Chtourou.
\newblock A roadmap for security challenges in the internet of things.
\newblock {\em Digital Communications and Networks}, 4(2):118--137, 2018.

\bibitem{Laprie:2005}
Jean-Claude Laprie.
\newblock Resilience for the scalability of dependability.
\newblock In {\em Network Computing and Applications, fourth IEEE international
  symposium on}, pages 5--6. IEEE, 2005.

\bibitem{han2014intrusion}
Song Han, Miao Xie, Hsiao-Hwa Chen, and Yun Ling.
\newblock Intrusion detection in cyber-physical systems: Techniques and
  challenges.
\newblock {\em IEEE systems journal}, 8(4):1052--1062, 2014.

\bibitem{yaqoob2017internet}
I.~Yaqoob, E.~Ahmed, I.~A.~T. Hashem, A.~I.~A. Ahmed, A.~Gani, M.~Imran, and
  M.~Guizani.
\newblock Internet of {{Things Architecture}}: {{Recent Advances}},
  {{Taxonomy}}, {{Requirements}}, and {{Open Challenges}}.
\newblock {\em IEEE Wireless Communications}, 24(3):10--16, June 2017.

\bibitem{shafique2018intelligent}
Muhammad Shafique, Faiq Khalid, and Semeen Rehman.
\newblock Intelligent security measures for smart cyber physical systems.
\newblock In {\em 2018 21st Euromicro Conference on Digital System Design
  (DSD)}, pages 280--287. IEEE, 2018.

\bibitem{wurm2017introduction}
Jacob Wurm, Yier Jin, Yang Liu, Shiyan Hu, Kenneth Heffner, Fahim Rahman, and
  Mark Tehranipoor.
\newblock Introduction to cyber-physical system security: A cross-layer
  perspective.
\newblock {\em IEEE Trans. Multi-Scale Comput. Syst}, 3(3):215--227, 2017.

\bibitem{myers2017automated}
II~Myers et~al.
\newblock Automated security domain partitioning with a formal method
  perspective of a cyber-physical systems.
\newblock 2017.

\bibitem{makedon2009event}
Fillia Makedon, Zhengyi Le, Heng Huang, Eric Becker, and Dimitrios Kosmopoulos.
\newblock An event driven framework for assistive cps environments.
\newblock {\em ACM SIGBED Review}, 6(2):3, 2009.

\bibitem{zdancewic2001secure}
Steve Zdancewic and Andrew~C Myers.
\newblock Secure information flow and cps.
\newblock In {\em European Symposium on Programming}, pages 46--61. Springer,
  2001.

\bibitem{xu2017security}
Qian Xu, Pinyi Ren, Houbing Song, and Qinghe Du.
\newblock Security-aware waveforms for enhancing wireless communications
  privacy in cyber-physical systems via multipath receptions.
\newblock {\em IEEE Internet of Things Journal}, 4(6):1924--1933, 2017.

\bibitem{chhetri2017fix}
Sujit~Rokka Chhetri, Sina Faezi, and Mohammad Abdullah~Al Faruque.
\newblock Fix the leak!: an information leakage aware secured cyber-physical
  manufacturing system.
\newblock In {\em Proceedings of the Conference on Design, Automation \& Test
  in Europe}, pages 1412--1417. European Design and Automation Association,
  2017.

\bibitem{conti2018leaky}
Mauro Conti.
\newblock Leaky cps: Inferring cyber information from physical properties (and
  the other way around).
\newblock In {\em Proceedings of the 4th ACM Workshop on Cyber-Physical System
  Security}, pages 23--24. ACM, 2018.

\bibitem{chhetri2018information}
Sujit~Rokka Chhetri, Sina Faezi, and Mohammad~Abdullah Al~Faruque.
\newblock Information leakage-aware computer-aided cyber-physical
  manufacturing.
\newblock {\em IEEE Transactions on Information Forensics and Security},
  13(9):2333--2344, 2018.

\bibitem{kriebel2018robustness}
Florian Kriebel, Semeen Rehman, Muhammad~Abdullah Hanif, Faiq Khalid, and
  Muhammad Shafique.
\newblock Robustness for smart cyber physical systems and internet-of-things:
  From adaptive robustness methods to reliability and security for machine
  learning.
\newblock In {\em 2018 IEEE Computer Society Annual Symposium on VLSI
  (ISVLSI)}, pages 581--586. IEEE, 2018.

\bibitem{balda2017cybersecurity}
Juan~Carlos Balda, Alan Mantooth, Rick Blum, and Paolo Tenti.
\newblock Cybersecurity and power electronics: Addressing the security
  vulnerabilities of the internet of things.
\newblock {\em IEEE Power Electronics Magazine}, 4(4):37--43, 2017.

\bibitem{mo2009secure}
Yilin Mo and Bruno Sinopoli.
\newblock Secure control against replay attacks.
\newblock In {\em Communication, Control, and Computing, 2009. Allerton 2009.
  47th Annual Allerton Conference on}, pages 911--918. IEEE, 2009.

\bibitem{ali2018wsn}
Saqib Ali, Taiseera Al~Balushi, Zia Nadir, and Omar~Khadeer Hussain.
\newblock Wsn security mechanisms for cps.
\newblock In {\em Cyber Security for Cyber Physical Systems}, pages 65--87.
  Springer, 2018.

\bibitem{shoukry2013non}
Yasser Shoukry, Paul Martin, Paulo Tabuada, and Mani Srivastava.
\newblock Non-invasive spoofing attacks for anti-lock braking systems.
\newblock In {\em International Workshop on Cryptographic Hardware and Embedded
  Systems}, pages 55--72. Springer, 2013.

\bibitem{radcliffe2011hacking}
Jerome Radcliffe.
\newblock Hacking medical devices for fun and insulin: Breaking the human scada
  system.
\newblock In {\em Black Hat Conference presentation slides}, volume 2011, 2011.

\bibitem{adepu2017waterjam}
Sridhar Adepu, Jay Prakash, and Aditya Mathur.
\newblock Waterjam: An experimental case study of jamming attacks on a water
  treatment system.
\newblock In {\em Software Quality, Reliability and Security Companion (QRS-C),
  2017 IEEE International Conference on}, pages 341--347. IEEE, 2017.

\bibitem{li2015jamming}
Yuzhe Li, Ling Shi, Peng Cheng, Jiming Chen, and Daniel~E Quevedo.
\newblock Jamming attacks on remote state estimation in cyber-physical systems:
  A game-theoretic approach.
\newblock {\em IEEE Transactions on Automatic Control}, 60(10):2831--2836,
  2015.

\bibitem{peng2018energy}
Lianghong Peng, Xianghui Cao, Changyin Sun, Yu~Cheng, and Shi Jin.
\newblock Energy efficient jamming attack schedule against remote state
  estimation in wireless cyber-physical systems.
\newblock {\em Neurocomputing}, 272:571--583, 2018.

\bibitem{peng2018optimal}
Lianghong Peng, Xianghui Cao, Hongbao Shi, and Changyin Sun.
\newblock Optimal jamming attack schedule for remote state estimation with two
  sensors.
\newblock {\em Journal of the Franklin Institute}, 2018.

\bibitem{zheng2017towards}
Zhiyuan Zheng and AL~Reddy.
\newblock Towards improving data validity of cyber-physical systems through
  path redundancy.
\newblock In {\em Proceedings of the 3rd ACM Workshop on Cyber-Physical System
  Security}, pages 91--102. ACM, 2017.

\bibitem{cao2017probabilistic}
Xianghui Cao and Changyin Sun.
\newblock Probabilistic denial of service attack against remote state
  estimation over a markov channel in cyber-physical systems.
\newblock In {\em Control Conference (ASCC), 2017 11th Asian}, pages 946--951.
  IEEE, 2017.

\bibitem{wang2018data}
Jun-Sheng Wang and Guang-Hong Yang.
\newblock Data-driven methods for stealthy attacks on tcp/ip-based networked
  control systems equipped with attack detectors.
\newblock {\em IEEE transactions on cybernetics}, (99):1--12, 2018.

\bibitem{dadras2018insider}
Soodeh Dadras and Chris Winstead.
\newblock Insider vs. outsider threats to autonomous vehicle platooning.
\newblock 2018.

\bibitem{hatzivasilis2017scotres}
George Hatzivasilis, Ioannis Papaefstathiou, and Charalampos Manifavas.
\newblock Scotres: Secure routing for iot and cps.
\newblock {\em IEEE Internet of Things Journal}, 4(6):2129--2141, 2017.

\bibitem{ali2018distributed}
Saqib Ali, Taiseera Al~Balushi, Zia Nadir, and Omar~Khadeer Hussain.
\newblock Distributed control systems security for cps.
\newblock In {\em Cyber Security for Cyber Physical Systems}, pages 141--160.
  Springer, 2018.

\bibitem{amrouch2017emerging}
Hussam Amrouch, Prashanth Krishnamurthy, Naman Patel, J{\"o}rg Henkel, Ramesh
  Karri, and Farshad Khorrami.
\newblock Emerging (un-) reliability based security threats and mitigations for
  embedded systems: special session.
\newblock In {\em Proceedings of the 2017 International Conference on
  Compilers, Architectures and Synthesis for Embedded Systems Companion},
  page~17. ACM, 2017.

\bibitem{poudel2017design}
Bikash Poudel, Naresh~Kumar Giri, and Arslan Munir.
\newblock Design and comparative evaluation of gpgpu-and fpga-based mpsoc ecu
  architectures for secure, dependable, and real-time automotive cps.
\newblock In {\em Application-specific Systems, Architectures and Processors
  (ASAP), 2017 IEEE 28th International Conference on}, pages 29--36. IEEE,
  2017.

\bibitem{lanotte2017formal}
Ruggero Lanotte, Massimo Merro, Riccardo Muradore, and Luca Vigan{\`o}.
\newblock A formal approach to cyber-physical attacks.
\newblock In {\em Computer Security Foundations Symposium (CSF), 2017 IEEE
  30th}, pages 436--450. IEEE, 2017.

\bibitem{yampolskiy2018security}
Mark Yampolskiy, Wayne~E King, Jacob Gatlin, Sofia Belikovetsky, Adam Brown,
  Anthony Skjellum, and Yuval Elovici.
\newblock Security of additive manufacturing: Attack taxonomy and survey.
\newblock {\em Additive Manufacturing}, 2018.

\bibitem{dong2017jamming}
Yanbo Dong and Peng Zhou.
\newblock Jamming attacks against control systems: A survey.
\newblock In {\em Intelligent Computing, Networked Control, and Their
  Engineering Applications}, pages 566--574. Springer, 2017.

\bibitem{nichols2018hybrid}
William~M Nichols.
\newblock {\em Hybrid Attack Graphs for Use with a Simulation of a
  Cyber-Physical System}.
\newblock PhD thesis, The University of Tulsa, 2018.

\bibitem{moore2017power}
Samuel~B Moore, Jacob Gatlin, Sofia Belikovetsky, Mark Yampolskiy, Wayne~E
  King, and Yuval Elovici.
\newblock Power consumption-based detection of sabotage attacks in additive
  manufacturing.
\newblock {\em arXiv preprint arXiv:1709.01822}, 2017.

\bibitem{rekhis2017securing}
S~Rekhis, N~Boudriga, and N~Ellouze.
\newblock Securing implantable medical devices against cyberspace attacks.
\newblock In {\em Anti-Cyber Crimes (ICACC), 2017 2nd International Conference
  on}, pages 187--192. IEEE, 2017.

\bibitem{aguma2018introduction}
Julius~C Aguma and Bruce McMillin.
\newblock Introduction of a hybrid monitor to cyber-physical systems.
\newblock {\em arXiv preprint arXiv:1805.01975}, 2018.

\bibitem{ahmed2017noisense}
Chuadhry~Mujeeb Ahmed, Aditya Mathur, and Martin Ochoa.
\newblock {\em arXiv preprint arXiv:1712.01598}, 2017.

\bibitem{gunduz2018reliability}
Hasan Gunduz and Dilan Jayaweera.
\newblock Reliability assessment of a power system with cyber-physical
  interactive operation of photovoltaic systems.
\newblock {\em International Journal of Electrical Power \& Energy Systems},
  101:371--384, 2018.

\bibitem{fink2017security}
Glenn~A Fink, Thomas~W Edgar, Theora~R Rice, Douglas~G MacDonald, and Cary~E
  Crawford.
\newblock Security and privacy in cyber-physical systems.
\newblock In {\em Cyber-Physical Systems}, pages 129--141. Elsevier, 2017.

\bibitem{marquis2018toward}
Victoria Marquis, Rebecca Ho, William Rainey, Matthew Kimpel, Joseph Ghiorzi,
  William Cricchi, and Nicola Bezzo.
\newblock Toward attack-resilient state estimation and control of autonomous
  cyber-physical systems.
\newblock In {\em Systems and Information Engineering Design Symposium (SIEDS),
  2018}, pages 70--75. IEEE, 2018.

\bibitem{Kephart:2003}
J.~O. Kephart and D.~M. Chess.
\newblock The vision of autonomic computing.
\newblock {\em Computer}, 36(1):41--50, Jan 2003.

\bibitem{Henzinger96}
Thomas~A. Henzinger.
\newblock The theory of hybrid automata.
\newblock In {\em Proceedings, 11th Annual {IEEE} Symposium on Logic in
  Computer Science, New Brunswick, New Jersey, USA, July 27-30, 1996}, pages
  278--292. {IEEE} Computer Society, 1996.

\bibitem{Bissmeyer:2012}
N.~Bi{\ss}meyer, S.~Mauthofer, K.~M. Bayarou, and F.~Kargl.
\newblock {Assessment of node trustworthiness in VANETs using data plausibility
  checks with particle filters}.
\newblock In {\em 2012 IEEE Vehicular Networking Conference (VNC)}, pages
  78--85, Nov 2012.

\bibitem{fauri2017system}
Davide Fauri, Daniel~Ricardo dos Santos, Elisa Costante, Jerry den Hartog,
  Sandro Etalle, and Stefano Tonetta.
\newblock From system specification to anomaly detection (and back).
\newblock In {\em Proceedings of the 2017 Workshop on Cyber-Physical Systems
  Security and PrivaCy}, pages 13--24. ACM, 2017.

\bibitem{Castro:2002}
Miguel Castro and Barbara Liskov.
\newblock Practical byzantine fault tolerance and proactive recovery.
\newblock {\em ACM Transactions on Computer Systems (TOCS)}, 20(4):398--461,
  2002.

\bibitem{Falcone10}
Yli\`{e}s Falcone.
\newblock You should better enforce than verify.
\newblock In Howard Barringer, Yli\`{e}s Falcone, Bernd Finkbeiner, Klaus
  Havelund, Insup Lee, Gordon~J. Pace, Grigore Rosu, Oleg Sokolsky, and Nikolai
  Tillmann, editors, {\em Proceedings of the 1st international conference on
  Runtime verification (RV 2010)}, volume 6418 of {\em Lecture Notes in
  Computer Science}, pages 89--105. Springer-Verlag, 2010.

\bibitem{Meng:2018}
W.~Meng, E.~W. Tischhauser, Q.~Wang, Y.~Wang, and J.~Han.
\newblock {When Intrusion Detection Meets Blockchain Technology: A Review}.
\newblock {\em IEEE Access}, 6:10179--10188, 2018.

\bibitem{FalconeMRS18}
Yli{\`{e}}s Falcone, Leonardo Mariani, Antoine Rollet, and Saikat Saha.
\newblock Runtime failure prevention and reaction.
\newblock In {\em Lectures on Runtime Verification - Introductory and Advanced
  Topics}, volume 10457, pages 103--134. Springer, 2018.

\bibitem{Petit:2015}
J.~Petit and S.~E. Shladover.
\newblock {Potential Cyberattacks on Automated Vehicles}.
\newblock {\em IEEE Transactions on Intelligent Transportation Systems},
  16(2):546--556, April 2015.

\bibitem{Poledna:1996}
Stefan Poledna.
\newblock {\em {Fault-Tolerant Real-Time Systems: The Problem of Replica
  Determinism}}.
\newblock Kluwer Academic Publishers, Norwell, MA, USA, 1996.

\bibitem{kriebel2016variability}
Florian Kriebel, Muhammad Shafique, Semeen Rehman, J{\"o}rg Henkel, and
  Siddharth Garg.
\newblock Variability and reliability awareness in the age of dark silicon.
\newblock {\em IEEE Design \& Test}, 33(2):59--67, 2016.

\bibitem{Duerrwang:2017}
Jürgen Dürrwang, Marcel Rumez, Johannes Braun, and Reiner Kriesten.
\newblock {Security Hardening with Plausibility Checks for Automotive ECUs}.
\newblock In {\em VEHICULAR 2017: The Sixth International Conference on
  Advances in Vehicular Systems, Technologies and Applications}, pages 38--41.
  IARIA, 07 2017.

\bibitem{Althoff2013}
Matthias Althoff.
\newblock Reachability analysis of nonlinear systems using conservative
  polynomialization and non-convex sets.
\newblock In {\em Proc. of HSCC '13: the 16th International Conference on
  Hybrid Systems: Computation and Control}, pages 173--182. ACM, 2013.

\bibitem{Franzle2007}
Martin Fr{\"a}nzle and Christian Herde.
\newblock Hysat: An efficient proof engine for bounded model checking of hybrid
  systems.
\newblock {\em Formal Methods in System Design}, 30(3):179--198, 2007.

\bibitem{AsarinDM02}
Eugene Asarin, Thao Dang, and Oded Maler.
\newblock The d/dt tool for verification of hybrid systems.
\newblock In {\em Computer Aided Verification, 14th International Conference,
  {CAV} 2002,Copenhagen, Denmark, July 27-31, 2002, Proceedings}, volume 2404
  of {\em LNCS}, pages 365--370. Springer, 2002.

\bibitem{RayGDBBG15}
Rajarshi Ray, Amit Gurung, Binayak Das, Ezio Bartocci, Sergiy Bogomolov, and
  Radu Grosu.
\newblock Xspeed: Accelerating reachability analysis on multi-core processors.
\newblock In {\em HVC}, volume 9434 of {\em LNCS}, pages 3--18. Springer, 2015.

\bibitem{chen2013flow}
Xin Chen, Erika {\'A}brah{\'a}m, and Sriram Sankaranarayanan.
\newblock Flow*: An analyzer for non-linear hybrid systems.
\newblock In {\em CAV}, pages 258--263, 2013.

\bibitem{FrehseLGDCRLRGDM11}
Goran Frehse, Colas {Le~{G}uernic}, Alexandre Donz\'e, Scott Cotton, Rajarshi
  Ray, Olivier Lebeltel, Rodolfo Ripado, Antoine Girard, Thao Dang, and Oded
  Maler.
\newblock Space{E}x: Scalable verification of hybrid systems.
\newblock In Shaz~Qadeer Ganesh~Gopalakrishnan, editor, {\em CAV}, LNCS.
  Springer, 2011.

\bibitem{kong2015dreach}
Soonho Kong, Sicun Gao, Wei Chen, and Edmund Clarke.
\newblock d{R}each: $\delta$-reachability analysis for hybrid systems.
\newblock In {\em TACAS}, pages 200--205, 2015.

\bibitem{duggirala2015c2e2}
Parasara~Sridhar Duggirala, Sayan Mitra, Mahesh Viswanathan, and Matthew Potok.
\newblock C2{E}2: {A} verification tool for stateflow models.
\newblock In {\em TACAS}, pages 68--82. Springer, 2015.

\bibitem{KongBH18}
Hui Kong, Ezio Bartocci, and Thomas~A. Henzinger.
\newblock Reachable set over-approximation for nonlinear systems using
  piecewise barrier tubes.
\newblock In {\em Proc. of {CAV} 2018: the 30th International Conference on
  Computer Aided Verification}, volume 10981 of {\em LNCS}, pages 449--467.
  Springer, 2018.

\bibitem{BartocciFFR18}
Ezio Bartocci, Yli{\`{e}}s Falcone, Adrian Francalanza, and Giles Reger.
\newblock Introduction to runtime verification.
\newblock In {\em Lectures on Runtime Verification - Introductory and Advanced
  Topics}, volume 10457 of {\em Lecture Notes in Computer Science}, pages
  1--33. Springer, 2018.

\bibitem{fpga}
Stefan Jaksic, Ezio Bartocci, Radu Grosu, Reinhard Kloibhofer, Thang Nguyen,
  and Dejan Ni\v{c}kovi\'{c}.
\newblock From signal temporal logic to {FPGA} monitors.
\newblock In {\em Proc. of {MEMOCODE} 2015: the 13th {ACM/IEEE} International
  Conference on Formal Methods and Models for Codesign}, pages 218--227.
  {IEEE}, 2015.

\bibitem{SelyuninJNRHBNG17}
Konstantin Selyunin, Stefan Jaksic, Thang Nguyen, Christian Reidl, Udo Hafner,
  Ezio Bartocci, Dejan Nickovic, and Radu Grosu.
\newblock Runtime monitoring with recovery of the {SENT} communication
  protocol.
\newblock In {\em Proc. of {CAV} 2017: the 29th International Conference on
  Computer Aided Verification}, volume 10426 of {\em LNCS}, pages 336--355.
  Springer, 2017.

\bibitem{ltl}
Amir Pnueli.
\newblock The temporal logic of programs.
\newblock In {\em Proc. of the 18th Annual Symposium on Foundations of Computer
  Science}, pages 46--57. IEEE, 1977.

\bibitem{Alur1994}
Rajeev Alur and Thomas~A. Henzinger.
\newblock A really temporal logic.
\newblock {\em Journal of {ACM}}, 41(1):181--204, 1994.

\bibitem{Maler2004}
Oded Maler and Dejan Nickovic.
\newblock Monitoring {T}emporal {P}roperties of {C}ontinuous {S}ignals.
\newblock In {\em Proc. of FORMATS-FTRTFT 2004: the Joint International
  Conferences on Formal Modeling and Analysis of Timed Systmes and Formal
  Techniques in Real-Time and Fault-Tolerant Systems}, volume 3253 of {\em
  LNCS}, pages 152--166. Springer-Verlag, 2004.

\bibitem{Donze2012}
Alexandre Donz{\'{e}}, Oded Maler, Ezio Bartocci, Dejan Ni\v{c}kovi\'{c}, Radu
  Grosu, and Scott~A. Smolka.
\newblock On temporal logic and signal processing.
\newblock In {\em Proc. of {ATVA} 2012: the 10th International Symposium on
  Automated Technology for Verification and Analysis}, volume 7561 of {\em
  LNCS}, pages 92--106. Springer, 2012.

\bibitem{jones2014anomaly}
Austin Jones, Zhaodan Kong, and Calin Belta.
\newblock Anomaly detection in cyber-physical systems: A formal methods
  approach.
\newblock In {\em Decision and Control (CDC), 2014 IEEE 53rd Annual Conference
  on}, pages 848--853. IEEE, 2014.

\bibitem{Clarkson2010}
Michael~R. Clarkson and Fred~B. Schneider.
\newblock Hyperproperties.
\newblock {\em J. Comput. Secur.}, 18(6):1157--1210, 2010.

\bibitem{ClarksonFKMRS14}
Michael~R. Clarkson, Bernd Finkbeiner, Masoud Koleini, Kristopher~K. Micinski,
  Markus~N. Rabe, and C{\'{e}}sar S{\'{a}}nchez.
\newblock Temporal logics for hyperproperties.
\newblock In {\em Proc. of {POST} 2014: the Third International Conference on
  Principles of Security and Trust}, volume 8414 of {\em LNCS}, pages 265--284.
  Springer, 2014.

\bibitem{Nguyen2017hyper}
Luan~Viet Nguyen, James Kapinski, Xiaoqing Jin, Jyotirmoy~V. Deshmukh, and
  Taylor~T. Johnson.
\newblock Hyperproperties of real-valued signals.
\newblock In {\em Proc. of MEMOCODE '17: the 15th ACM-IEEE International
  Conference on Formal Methods and Models for System Design}, MEMOCODE '17,
  pages 104--113. ACM, 2017.

\bibitem{BonakdarpourF16}
Borzoo Bonakdarpour and Bernd Finkbeiner.
\newblock Runtime verification for hyperltl.
\newblock In {\em Proc. of {RV} 2016: the 16th International Conference on
  Runtime Verification}, volume 10012 of {\em LNCS}, pages 41--45. Springer,
  2016.

\bibitem{fainekos-robust}
Georgios~E. Fainekos and George~J. Pappas.
\newblock Robustness of temporal logic specifications for continuous-time
  signals.
\newblock {\em Theor. Comput. Sci.}, 410(42):4262--4291, 2009.

\bibitem{robust1}
Alexandre Donz{\'{e}} and Oded Maler.
\newblock Robust satisfaction of temporal logic over real-valued signals.
\newblock In {\em Proc. of {FORMATS}'10: the 8th International Conference on
  Formal Modeling and Analysis of Timed Systems}, volume 6246 of {\em LNCS},
  pages 92--106. Springer, 2010.

\bibitem{staliro}
Yashwanth Annpureddy, Che Liu, Georgios~E. Fainekos, and Sriram
  Sankaranarayanan.
\newblock {S}-{T}a{L}i{R}o: {A} tool for temporal logic falsification for
  hybrid systems.
\newblock In {\em Proc. of {TACAS} 2011: the 17th International Conference on
  Tools and Algorithms for the Construction and Analysis of Systems}, volume
  6605 of {\em LNCS}, pages 254--257, 2011.

\bibitem{breach}
Alexandre Donz{\'{e}}.
\newblock Breach, {A} toolbox for verification and parameter synthesis of
  hybrid systems.
\newblock In {\em Proc. of {CAV} 2010: the 22nd International Conference on
  Computer Aided Verification}, volume 6174 of {\em LNCS}, pages 167--170.
  Springer, 2010.

\bibitem{DokhanchiEtal2015emsoft}
Adel Dokhanchi, Aditya Zutshi, Rahul~T. Sriniva, Sriram Sankaranarayanan, and
  Georgios Fainekos.
\newblock Requirements driven falsification with coverage metrics.
\newblock In {\em Proc. of EMSOFT: the 12th International Conference on
  Embedded Software}, pages 31--40. IEEE, 2015.

\bibitem{DonzeKR09hscc}
Alexandre Donz{\'{e}}, Bruce Krogh, and Akshay Rajhans.
\newblock Parameter synthesis for hybrid systems with an application to
  simulink models.
\newblock In {\em Proc. of HSCC 2009: the 12th International Conference on
  Hybrid Systems: Computation and Control}, volume 5469 of {\em LNCS}, pages
  165--179. Springer, 2009.

\bibitem{BartocciBNS15}
Ezio Bartocci, Luca Bortolussi, Laura Nenzi, and Guido Sanguinetti.
\newblock System design of stochastic models using robustness of temporal
  properties.
\newblock {\em Theor. Comput. Sci.}, 587:3--25, 2015.

\bibitem{David:2015}
O.~E. David and N.~S. Netanyahu.
\newblock Deepsign: Deep learning for automatic malware signature generation
  and classification.
\newblock In {\em 2015 International Joint Conference on Neural Networks
  (IJCNN)}, pages 1--8, July 2015.

\bibitem{Wilcoxon:1945}
Frank Wilcoxon.
\newblock Individual comparisons by ranking methods.
\newblock {\em Biometrics Bulletin}, 1(6):80--83, 1945.

\bibitem{Malhotra:2015}
Pankaj Malhotra, Lovekesh Vig, Gautam Shroff, and Puneet Agarwal.
\newblock {Long Short Term Memory Networks for Anomaly Detection in Time
  Series}.
\newblock In {\em European Symposium on Artificial Neural Networks,
  Computational Intelligence and Machine Learning (ESANN)}, pages 89--94, April
  2015.

\bibitem{Lee:2001}
Wenke Lee and Dong Xiang.
\newblock Information-theoretic measures for anomaly detection.
\newblock In {\em Security and Privacy, 2001. S\&P 2001. Proceedings. 2001 IEEE
  Symposium on}, pages 130--143. IEEE, 2001.

\bibitem{Wilson:1993}
Paul~F Wilson.
\newblock {\em Root cause analysis: A tool for total quality management}.
\newblock ASQ Quality Press, 1993.

\bibitem{BartocciFMN18}
Ezio Bartocci, Thomas Ferr{\`{e}}re, Niveditha Manjunath, and Dejan Nickovic.
\newblock Localizing faults in simulink/stateflow models with {STL}.
\newblock In {\em Proc. of {HSCC} 2018: the 21st International Conference on
  Hybrid Systems: Computation and Control}, pages 197--206, 2018.

\bibitem{LiuLNBB16}
Bing Liu, Lucia, Shiva Nejati, Lionel~C. Briand, and Thomas Bruckmann.
\newblock Localizing multiple faults in simulink models.
\newblock In {\em International Conference on Software Analysis, Evolution, and
  Reengineering}, pages 146--156. {IEEE} Computer Society, 2016.

\bibitem{LiuLNBB16b}
Bing Liu, Lucia, Shiva Nejati, Lionel~C. Briand, and Thomas Bruckmann.
\newblock Simulink fault localization: an iterative statistical debugging
  approach.
\newblock {\em Softw. Test., Verif. Reliab.}, 26(6):431--459, 2016.

\bibitem{LiuLNB17}
Bing Liu, Lucia, Shiva Nejati, and Lionel~C. Briand.
\newblock Improving fault localization for simulink models using search-based
  testing and prediction models.
\newblock In {\em International Conference on Software Analysis, Evolution and
  Reengineering}, pages 359--370. {IEEE} Computer Society, 2017.

\bibitem{4344104}
R.~Abreu, P.~Zoeteweij, and A.~J.~C. van Gemund.
\newblock On the accuracy of spectrum-based fault localization.
\newblock In {\em Testing: Academic and Industrial Conference Practice and
  Research Techniques}, pages 89--98. IEEE, 2007.

\bibitem{DeshmukhJMP18}
Jyotirmoy~V. Deshmukh, Xiaoqing Jin, Rupak Majumdar, and Vinayak~S. Prabhu.
\newblock Parameter optimization in control software using statistical fault
  localization techniques.
\newblock In {\em Proc. of {ICCPS} 2018: the 9th {ACM/IEEE} International
  Conference on Cyber-Physical Systems}, pages 220--231. {IEEE} / {ACM}, 2018.

\bibitem{WongGLAW16}
W.~Eric Wong, Ruizhi Gao, Yihao Li, Rui Abreu, and Franz Wotawa.
\newblock A survey on software fault localization.
\newblock {\em {IEEE} Trans. Software Eng.}, 42(8):707--740, 2016.

\bibitem{FerrereMalerNickovic15}
Thomas Ferr{\`{e}}re, Oded Maler, and Dejan Nickovic.
\newblock Trace diagnostics using temporal implicants.
\newblock In {\em International Symposium on Automated Technology for
  Verification and Analysis}, volume 9364 of {\em LNCS}, pages 241--258.
  Springer, 2015.

\bibitem{Marler:2004}
R~Timothy Marler and Jasbir~S Arora.
\newblock Survey of multi-objective optimization methods for engineering.
\newblock {\em Structural and multidisciplinary optimization}, 26(6):369--395,
  2004.

\bibitem{Mehra:1970}
R.~Mehra.
\newblock On the identification of variances and adaptive kalman filtering.
\newblock {\em IEEE Transactions on Automatic Control}, 15(2):175--184, April
  1970.

\bibitem{Schneider00}
Fred~B. Schneider.
\newblock Enforceable security policies.
\newblock {\em {ACM} Trans. Inf. Syst. Secur.}, 3(1):30--50, 2000.

\bibitem{LigattiBW05}
Jay Ligatti, Lujo Bauer, and David Walker.
\newblock Edit automata: enforcement mechanisms for run-time security policies.
\newblock {\em Int. J. Inf. Sec.}, 4(1-2):2--16, 2005.

\bibitem{FalconeMFR11}
Yli{\`{e}}s Falcone, Laurent Mounier, Jean{-}Claude Fernandez, and Jean{-}Luc
  Richier.
\newblock Runtime enforcement monitors: composition, synthesis, and enforcement
  abilities.
\newblock {\em Formal Methods in System Design}, 38(3):223--262, 2011.

\bibitem{BielovaM12}
Nataliia Bielova and Fabio Massacci.
\newblock Iterative enforcement by suppression: Towards practical enforcement
  theories.
\newblock {\em Journal of Computer Security}, 20(1):51--79, 2012.

\bibitem{DolzhenkoLR15}
Egor Dolzhenko, Jay Ligatti, and Srikar Reddy.
\newblock Modeling runtime enforcement with mandatory results automata.
\newblock {\em Int. J. Inf. Sec.}, 14(1):47--60, 2015.

\bibitem{Ylies2016}
Yli{\`{e}}s Falcone, Thierry J{\'{e}}ron, Herv{\'{e}} Marchand, and Srinivas
  Pinisetty.
\newblock Runtime enforcement of regular timed properties by suppressing and
  delaying events.
\newblock {\em Systems {\&} Control Letters}, 123:2--41, 2016.

\bibitem{PinisettyFJMRN14}
Srinivas Pinisetty, Yli{\`{e}}s Falcone, Thierry J{\'{e}}ron, Herv{\'{e}}
  Marchand, Antoine Rollet, and Omer Nguena{-}Timo.
\newblock Runtime enforcement of timed properties revisited.
\newblock {\em Formal Methods in System Design}, 45(3):381--422, 2014.

\bibitem{FalconeM15}
Yli{\`{e}}s Falcone and Herv{\'{e}} Marchand.
\newblock Enforcement and validation (at runtime) of various notions of
  opacity.
\newblock {\em Discrete Event Dynamic Systems}, 25(4):531--570, 2015.

\bibitem{RenardRF17}
Matthieu Renard, Antoine Rollet, and Yli{\`{e}}s Falcone.
\newblock Runtime enforcement using b{\"{u}}chi games.
\newblock In {\em Proceedings of the 24th {ACM} {SIGSOFT} International {SPIN}
  Symposium on Model Checking of Software}, pages 70--79. {ACM}, 2017.

\bibitem{Meneses:2015}
E.~Meneses, X.~Ni, G.~Zheng, C.~L. Mendes, and L.~V. Kalé.
\newblock {Using Migratable Objects to Enhance Fault Tolerance Schemes in
  Supercomputers}.
\newblock {\em IEEE Transactions on Parallel and Distributed Systems},
  26(7):2061--2074, July 2015.

\bibitem{Sancho:2005}
J.~C. Sancho, F.~Petrini, K.~Davis, R.~Gioiosa, and S.~Jiang.
\newblock Current practice and a direction forward in checkpoint/restart
  implementations for fault tolerance.
\newblock In {\em 19th IEEE International Parallel and Distributed Processing
  Symposium}, April 2005.

\bibitem{Ghosh:1997}
S.~Ghosh, R.~Melhem, and D.~Mosse.
\newblock Fault-tolerance through scheduling of aperiodic tasks in hard
  real-time multiprocessor systems.
\newblock {\em IEEE Transactions on Parallel and Distributed Systems},
  8(3):272--284, Mar 1997.

\bibitem{li2013raster}
Tuo Li, Muhammad Shafique, Jude~Angelo Ambrose, Semeen Rehman, J{\"o}rg Henkel,
  and Sri Parameswaran.
\newblock Raster: Runtime adaptive spatial/temporal error resiliency for
  embedded processors.
\newblock In {\em Design Automation Conference (DAC), 2013 50th ACM/EDAC/IEEE},
  pages 1--7. IEEE, 2013.

\bibitem{li2013dhaser}
Tuo Li, Muhammad Shafique, Semeen Rehman, Jude~Angelo Ambrose, J{\"o}rg Henkel,
  and Sri Parameswaran.
\newblock Dhaser: dynamic heterogeneous adaptation for soft-error resiliency in
  asip-based multi-core systems.
\newblock In {\em Computer-Aided Design (ICCAD), 2013 IEEE/ACM International
  Conference on}, pages 646--653. IEEE, 2013.

\bibitem{li2013cser}
Tuo Li, Muhammad Shafique, Semeen Rehman, Swarnalatha Radhakrishnan, Roshan
  Ragel, Jude~Angelo Ambrose, J{\"o}rg Henkel, and Sri Parameswaran.
\newblock Cser: Hw/sw configurable soft-error resiliency for application
  specific instruction-set processors.
\newblock In {\em Proceedings of the Conference on Design, Automation and Test
  in Europe}, pages 707--712. EDA Consortium, 2013.

\bibitem{li2017fine}
Tuo Li, Muhammad Shafique, Jude~Angelo Ambrose, J{\"o}rg Henkel, and Sri
  Parameswaran.
\newblock Fine-grained checkpoint recovery for application-specific
  instruction-set processors.
\newblock {\em IEEE Transactions on Computers}, 66(4):647--660, 2017.

\bibitem{Seto98}
D.~Seto, B.~Krogh, L.~Sha, and A.~Chutinan.
\newblock The simplex architecture for safe online control system upgrades.
\newblock In {\em Proc. of ACC 1998: the American Control Conference},
  volume~6, pages 3504--3508 vol.6, 1998.

\bibitem{shafique2014dark}
Muhammad Shafique, Siddharth Garg, Tulika Mitra, Sri Parameswaran, and J{\"o}rg
  Henkel.
\newblock Dark silicon as a challenge for hardware/software co-design: Invited
  special session paper.
\newblock In {\em Proceedings of the 2014 International Conference on
  Hardware/Software Codesign and System Synthesis}, page~13. ACM, 2014.

\bibitem{henkel2014multi}
Jorg Henkel, Lars Bauer, Hongyan Zhang, Semeen Rehman, and Muhammad Shafique.
\newblock Multi-layer dependability: From microarchitecture to application
  level.
\newblock In {\em Design Automation Conference (DAC), 2014 51st ACM/EDAC/IEEE},
  pages 1--6. IEEE, 2014.

\bibitem{henkel2013reliable}
J{\"o}rg Henkel, Lars Bauer, Nikil Dutt, Puneet Gupta, Sani Nassif, Muhammad
  Shafique, Mehdi Tahoori, and Norbert Wehn.
\newblock Reliable on-chip systems in the nano-era: Lessons learnt and future
  trends.
\newblock In {\em Proceedings of the 50th Annual Design Automation Conference},
  page~99. ACM, 2013.

\bibitem{rehman2014dtune}
Semeen Rehman, Florian Kriebel, Duo Sun, Muhammad Shafique, and J{\"o}rg
  Henkel.
\newblock dtune: Leveraging reliable code generation for adaptive dependability
  tuning under process variation and aging-induced effects.
\newblock In {\em Proceedings of the 51st Annual Design Automation Conference},
  pages 1--6. ACM, 2014.

\bibitem{rehman2011reliable}
Semeen Rehman, Muhammad Shafique, Florian Kriebel, and J{\"o}rg Henkel.
\newblock Reliable software for unreliable hardware: embedded code generation
  aiming at reliability.
\newblock In {\em Proceedings of the seventh IEEE/ACM/IFIP international
  conference on Hardware/software codesign and system synthesis}, pages
  237--246. ACM, 2011.

\bibitem{rehman2014reliability}
Semeen Rehman, Florian Kriebel, Muhammad Shafique, and Joerg Henkel.
\newblock Reliability-driven software transformations for unreliable hardware.
\newblock {\em IEEE Transactions on Computer-Aided Design of Integrated
  Circuits and Systems}, 33(11):1597--1610, 2014.

\bibitem{kriebel2014aser}
Florian Kriebel, Semeen Rehman, Duo Sun, Muhammad Shafique, and J{\"o}rg
  Henkel.
\newblock Aser: Adaptive soft error resilience for reliability-heterogeneous
  processors in the dark silicon era.
\newblock In {\em Proceedings of the 51st annual design automation conference},
  pages 1--6. ACM, 2014.

\bibitem{shafique2013exploiting}
Muhammad Shafique, Semeen Rehman, Pau~Vilimelis Aceituno, and J{\"o}rg Henkel.
\newblock Exploiting program-level masking and error propagation for
  constrained reliability optimization.
\newblock In {\em Proceedings of the 50th Annual Design Automation Conference},
  page~17. ACM, 2013.

\bibitem{gnad2015hayat}
Dennis Gnad, Muhammad Shafique, Florian Kriebel, Semeen Rehman, Duo Sun, and
  J{\"o}rg Henkel.
\newblock Hayat: Harnessing dark silicon and variability for aging deceleration
  and balancing.
\newblock In {\em Design Automation Conference (DAC), 2015 52nd ACM/EDAC/IEEE},
  pages 1--6. IEEE, 2015.

\bibitem{Hoeftberger:2015}
Oliver H{\"o}ftberger.
\newblock {\em {Knowledge-based Dynamic Reconfiguration for Embedded Real-Rime
  Systems}}.
\newblock PhD thesis, Technische Universit\"at Wien, 2015.

\bibitem{Ratasich:2018}
D.~Ratasich, T.~Preindl, K.~Selyunin, and R.~Grosu.
\newblock {Self-Healing by Property-Guided Structural Adaptation}.
\newblock In {\em 2018 IEEE 1st International Conference on Industrial
  Cyber-Physical Systems (ICPS)}, pages 199--205, May 2018.

\bibitem{Nakamoto:2008}
Satoshi Nakamoto.
\newblock Bitcoin: A peer-to-peer electronic cash system.
\newblock 2008.

\bibitem{Miller:2014}
Andrew Miller and Joseph~J LaViola~Jr.
\newblock Anonymous byzantine consensus from moderately-hard puzzles: A model
  for bitcoin.
\newblock {\em Available on line: http://nakamotoinstitute.
  org/research/anonymous-byzantine-consensus}, 2014.

\bibitem{Dorri:2017}
Ali Dorri, Salil~S Kanhere, Raja Jurdak, and Praveen Gauravaram.
\newblock Blockchain for iot security and privacy: The case study of a smart
  home.
\newblock In {\em Pervasive Computing and Communications Workshops (PerCom
  Workshops), 2017 IEEE International Conference on}, pages 618--623. IEEE,
  2017.

\bibitem{alphand2018iotchain}
Olivier Alphand, Michele Amoretti, Timothy Claeys, Simone Dall'Asta, Andrzej
  Duda, Gianluigi Ferrari, Franck Rousseau, Bernard Tourancheau, Luca Veltri,
  and Francesco Zanichelli.
\newblock Iotchain: A blockchain security architecture for the internet of
  things.
\newblock In {\em Wireless Communications and Networking Conference (WCNC),
  2018 IEEE}, pages 1--6. IEEE, 2018.

\bibitem{Alcaraz:2016}
Cristina Alcaraz and Javier Lopez.
\newblock {Safeguarding Structural Controllability in Cyber-Physical Control
  Systems}.
\newblock In Ioannis Askoxylakis, Sotiris Ioannidis, Sokratis Katsikas, and
  Catherine Meadows, editors, {\em Computer Security -- ESORICS 2016}, pages
  471--489, Cham, 2016. Springer International Publishing.

\bibitem{Khamphanchai:2011}
W.~Khamphanchai, S.~Pisanupoj, W.~Ongsakul, and M.~Pipattanasomporn.
\newblock A multi-agent based power system restoration approach in distributed
  smart grid.
\newblock In {\em 2011 International Conference Utility Exhibition on Power and
  Energy Systems: Issues and Prospects for Asia (ICUE)}, pages 1--7, Sept 2011.

\bibitem{Yan:2016}
Y.~Yan, B.~Zhang, and J.~Guo.
\newblock {An Adaptive Decision Making Approach Based on Reinforcement Learning
  for Self-Managed Cloud Applications}.
\newblock In {\em 2016 IEEE International Conference on Web Services (ICWS)},
  pages 720--723, June 2016.

\bibitem{Dragoni:2017microservices}
Nicola Dragoni, Saverio Giallorenzo, Alberto~Lluch Lafuente, Manuel Mazzara,
  Fabrizio Montesi, Ruslan Mustafin, and Larisa Safina.
\newblock Microservices: yesterday, today, and tomorrow.
\newblock In {\em Present and Ulterior Software Engineering}, pages 195--216.
  Springer, 2017.

\bibitem{bohrer2018veriphy}
Brandon Bohrer, Yong~Kiam Tan, Stefan Mitsch, Magnus~O Myreen, and Andr{\'e}
  Platzer.
\newblock Veriphy: verified controller executables from verified cyber-physical
  system models.
\newblock In {\em Proceedings of the 39th ACM SIGPLAN Conference on Programming
  Language Design and Implementation}, pages 617--630. ACM, 2018.

\bibitem{mohsin2016iotsat}
Mujahid Mohsin, Zahid Anwar, Ghaith Husari, Ehab Al-Shaer, and Mohammad~Ashiqur
  Rahman.
\newblock Iotsat: A formal framework for security analysis of the internet of
  things (iot).
\newblock In {\em Communications and Network Security (CNS), 2016 IEEE
  Conference on}, pages 180--188. IEEE, 2016.

\bibitem{mohsin2017iotriskanalyzer}
Mujahid Mohsin, Muhammad~Usama Sardar, Osman Hasan, and Zahid Anwar.
\newblock Iotriskanalyzer: A probabilistic model checking based framework for
  formal risk analytics of the internet of things.
\newblock {\em IEEE Access}, 5:5494--5505, 2017.

\bibitem{kang2018model}
Eun-Young Kang, Dongrui Mu, Li~Huang, and Qianqing Lan.
\newblock Model-based verification and validation of an autonomous vehicle
  system.
\newblock {\em arXiv preprint arXiv:1803.06103}, 2018.

\bibitem{munir2018design}
Arslan Munir and Farinaz Koushanfar.
\newblock Design and analysis of secure and dependable automotive cps: A
  steer-by-wire case study.
\newblock {\em IEEE Transactions on Dependable and Secure Computing}, 2018.

\bibitem{park2018study}
Seong-Taek Park, Guozhong Li, and Jae-Chang Hong.
\newblock A study on smart factory-based ambient intelligence context-aware
  intrusion detection system using machine learning.
\newblock {\em Journal of Ambient Intelligence and Humanized Computing}, pages
  1--8, 2018.

\bibitem{ching2018opportunities}
Travers Ching, Daniel~S Himmelstein, Brett~K Beaulieu-Jones, Alexandr~A
  Kalinin, Brian~T Do, Gregory~P Way, Enrico Ferrero, Paul-Michael Agapow,
  Michael Zietz, Michael~M Hoffman, et~al.
\newblock Opportunities and obstacles for deep learning in biology and
  medicine.
\newblock {\em Journal of The Royal Society Interface}, 15(141):20170387, 2018.

\bibitem{hasani2018re}
Ramin~M Hasani, Mathias Lechner, Alexander Amini, Daniela Rus, and Radu Grosu.
\newblock Re-purposing compact neuronal circuit policies to govern
  reinforcement learning tasks.
\newblock {\em arXiv preprint arXiv:1809.04423}, 2018.

\bibitem{hanif2018robust}
Muhammad~Abdullah Hanif, Faiq Khalid, Rachmad Vidya~Wicaksana Putra, Semeen
  Rehman, and Muhammad Shafique.
\newblock Robust machine learning systems: Reliability and security for deep
  neural networks.
\newblock In {\em 2018 IEEE 24th International Symposium on On-Line Testing And
  Robust System Design (IOLTS)}, pages 257--260. IEEE, 2018.

\bibitem{hanzlik2018mlcapsule}
Lucjan Hanzlik, Yang Zhang, Kathrin Grosse, Ahmed Salem, Max Augustin, Michael
  Backes, and Mario Fritz.
\newblock Mlcapsule: Guarded offline deployment of machine learning as a
  service.
\newblock {\em arXiv preprint arXiv:1808.00590}, 2018.

\bibitem{hynes2018efficient}
Nick Hynes, Raymond Cheng, and Dawn Song.
\newblock Efficient deep learning on multi-source private data.
\newblock {\em arXiv preprint arXiv:1807.06689}, 2018.

\bibitem{Riazi2018deep}
M.~Sadegh Riazi, Bita~Darvish Rouhani, and Farinaz Koushanfar.
\newblock Deep learning on private data.
\newblock {\em in IEEE Security and Privacy Magazine}, 2018.

\bibitem{chen2018deepmarks}
Huili Chen, Bita~Darvish Rohani, and Farinaz Koushanfar.
\newblock Deepmarks: A digital fingerprinting framework for deep neural
  networks.
\newblock {\em arXiv preprint arXiv:1804.03648}, 2018.

\bibitem{rouhani2018deepsigns}
Bita~Darvish Rouhani, Huili Chen, and Farinaz Koushanfar.
\newblock Deepsigns: A generic watermarking framework for ip protection of deep
  learning models.
\newblock {\em arXiv preprint arXiv:1804.00750}, 2018.

\bibitem{zantedeschi2017efficient}
Valentina Zantedeschi, Maria-Irina Nicolae, and Ambrish Rawat.
\newblock Efficient defenses against adversarial attacks.
\newblock In {\em Proceedings of the 10th ACM Workshop on Artificial
  Intelligence and Security}, pages 39--49. ACM, 2017.

\bibitem{rakin2018defend}
Adnan~Siraj Rakin, Jinfeng Yi, Boqing Gong, and Deliang Fan.
\newblock Defend deep neural networks against adversarial examples via fixed
  anddynamic quantized activation functions.
\newblock {\em arXiv preprint arXiv:1807.06714}, 2018.

\bibitem{papernot2015distillation}
Nicolas Papernot, Patrick McDaniel, Xi~Wu, Somesh Jha, and Ananthram Swami.
\newblock Distillation as a defense to adversarial perturbations against deep
  neural networks.
\newblock {\em arXiv preprint arXiv:1511.04508}, 2015.

\bibitem{xiang2018verification}
Weiming Xiang, Patrick Musau, Ayana~A Wild, Diego~Manzanas Lopez, Nathaniel
  Hamilton, Xiaodong Yang, Joel Rosenfeld, and Taylor~T Johnson.
\newblock Verification for machine learning, autonomy, and neural networks
  survey.
\newblock {\em arXiv preprint arXiv:1810.01989}, 2018.

\bibitem{xiang2018reachability}
Weiming Xiang and Taylor~T Johnson.
\newblock Reachability analysis and safety verification for neural network
  control systems.
\newblock {\em arXiv preprint arXiv:1805.09944}, 2018.

\bibitem{Ratasich:2017}
D.~Ratasich, O.~H{\"o}ftberger, H.~Isakovic, M.~Shafique, and R.~Grosu.
\newblock {A Self-Healing Framework for Building Resilient Cyber-Physical
  Systems}.
\newblock In {\em 2017 IEEE 20th International Symposium on Real-Time
  Distributed Computing (ISORC)}, pages 133--140, May 2017.

\bibitem{Kopetz:2014}
Hermann Kopetz.
\newblock {A Conceptual Model for the Information Transfer in
  Systems-of-Systems}.
\newblock In {\em 2014 IEEE 17th International Symposium on
  Object/Component/Service-Oriented Real-Time Distributed Computing}, pages
  17--24, June 2014.

\bibitem{0001SDKJ15}
Aditya Zutshi, Sriram Sankaranarayanan, Jyotirmoy~V. Deshmukh, James Kapinski,
  and Xiaoqing Jin.
\newblock Falsification of safety properties for closed loop control systems.
\newblock In {\em International Conference on Hybrid Systems: Computation and
  Control}, pages 299--300. {ACM}, 2015.

\bibitem{ReichertG12}
Robert Reicherdt and Sabine Glesner.
\newblock Slicing {MATLAB} simulink models.
\newblock In {\em International Conference on Software Engineering}, pages
  551--561. {IEEE} Computer Society, 2012.

\end{thebibliography}

\end{document}